\documentclass[journal]{IEEEtran}

\usepackage{cite}

\usepackage[pdftex]{graphicx}

\usepackage{amsmath}

\usepackage{amsfonts}
\usepackage{amsthm}

\theoremstyle{plain}
\newtheorem{thm}{Theorem}
\newtheorem{lem}{Lemma}
\newtheorem{prop}{Proposition}

\newtheorem{remark}{Remark}

\usepackage{algorithm}
\usepackage{algorithmic}
\usepackage{multirow}  
\usepackage{booktabs}  
\usepackage{bm}

\usepackage{array}

\usepackage{MnSymbol}

\usepackage{graphicx} 

\ifCLASSOPTIONcompsoc
\usepackage[caption=false,font=normalsize,labelfont=sf,textfont=sf]{subfig}
\else
\usepackage[caption=false,font=footnotesize]{subfig}
\fi

\usepackage{color}

\usepackage{stfloats}

\renewcommand{\arraystretch}{1.3}

\usepackage{nomencl}
\usepackage{ifthen}
\renewcommand{\nomgroup}[1]{%
	\ifthenelse{\equal{#1}{A}}{\item[\emph{\textbf{Set and Index}}]}{%
		\ifthenelse{\equal{#1}{B}}{\item[\emph{\textbf{Parameters}}]}{%
			\ifthenelse{\equal{#1}{C}}{\item[\emph{\textbf{Variables}}]}{%
				\ifthenelse{\equal{#1}{D}}{\item[\emph{\textbf{XX}}]}
			}
		}
	}
}
\makenomenclature

\begin{document}

\title{Robust Co-Optimization of Distribution Network Hardening and Mobile  Resource Scheduling with Decision-Dependent Uncertainty}

\author{Donglai Ma,
	Xiaoyu Cao,~\IEEEmembership{Member,~IEEE}, 
	Bo Zeng,~\IEEEmembership{Member,~IEEE},
	Chen Chen,~\IEEEmembership{Senior Member,~IEEE},	\\
	Qiaozhu Zhai,~\IEEEmembership{Member,~IEEE},	
	Qing-Shan Jia,~\IEEEmembership{Senior Member,~IEEE},
	and Xiaohong Guan,~\IEEEmembership{Life Fellow,~IEEE}
	
	\thanks{This work was partially supported by the National Key Research and Development Program of China under Grant 2022YFA1004600, and by the National Natural Science Foundation of China under Grant 62373294 and 62192752. \emph{(Corresponding author: Xiaoyu Cao.)}}%
	\thanks{Donglai Ma, Xiaoyu Cao and Qiaozhu Zhai are with the School of Automation Science and Engineering and the Ministry of Education Key Laboratory for Intelligent Networks and Network Security, Xi’an Jiaotong University, Xi’an 710049, Shaanxi, China (e-mail: mdl1622267691@stu.xjtu.edu.cn; cxykeven2019@xjtu.edu.cn; qzzhai@sei.xjtu.edu.cn).}%
	\thanks{Bo Zeng is with the Department of Industrial Engineering and the Department of Electrical and Computer Engineering, University of Pittsburgh, Pittsburgh, PA 15106 USA (e-mail: bzeng@pitt.edu).}%
	\thanks{Chen Chen is with the School of Electrical Engineering, Xi’an Jiaotong University, Xi’an 710049, Shaanxi, China (e-mail: morningchen@xjtu.edu.cn).}%
	\thanks{Qing-Shan Jia is with the Center for Intelligent and Networked Systems, Department of Automation, Tsinghua University, Beijing 100084, China (e-mail: jiaqs@mail.tsinghua.edu.cn).}%
	\thanks{Xiaohong Guan is with the School of Automation Science and Engineering and the Ministry of Education Key Laboratory for Intelligent Networks and Network Security, Xi’an Jiaotong University, Xi’an 710049, Shaanxi, China, and also with the Center for Intelligent and Networked Systems, Department of Automation, Tsinghua University, Beijing 100084, China (e-mail: xhguan@xjtu.edu.cn).}%
}

\maketitle

\begin{abstract}
	This paper studies the robust co-planning of proactive network hardening and mobile hydrogen energy resources (MHERs) scheduling, which is to enhance the resilience of power distribution network (PDN) against the disastrous events. A decision-dependent robust optimization model is formulated with $\min-\max$ resilience constraint and discrete recourse structure, which helps achieve the load survivability target considering endogenous uncertainties. Different from the traditional model with a fixed uncertainty set, we adopt a dynamic representation that explicitly captures the endogenous uncertainties of network contingency as well as the available hydrogen storage levels of MHERs, which induces a decision-dependent uncertainty (DDU) set. Also, the multi-period adaptive routing and energy scheduling of MHERs are modeled as a mixed-integer recourse problem for further decreasing the resilience cost. Then, a nested parametric column-and-constraint generation (N-PC\&CG) algorithm is customized and developed to solve this challenging formulation. By leveraging the structural property of the DDU set as well as the combination of discrete recourse decisions and the corresponding extreme points, we derive a strengthened solution scheme with nontrivial enhancement strategies to realize efficient and exact computation. Numerical results on 14-bus test system and 56-bus real-world distribution network demonstrate the resilience benefits and economical feasibility of the proposed method under different damage severity levels. Moreover, the enhanced N-PC\&CG shows a superior solution capability to support prompt decisions for resilient planning with DDU models.
\end{abstract}

\begin{IEEEkeywords}
Distribution network hardening, mobile hydrogen energy resources, decision-dependent robust optimization, resilience constraint, nested parametric C\&CG
\end{IEEEkeywords}

\IEEEpeerreviewmaketitle

\printnomenclature
\nomenclature[A]{$\mathcal{L}/(i,j)$}{Set/Index of lines in PDN}%
\nomenclature[A]{$\mathcal{M}/m$}{Set/Index of MHERs}%
\nomenclature[A]{$\mathcal{N}/j$}{Set/Index of nodes in PDN}%
\nomenclature[A]{$\mathcal{N}^{\rm{sub}}$}{Set of substation nodes}%
\nomenclature[A]{$\mathcal{N}^{eh}$}{Set of nodes with electricity-hydrogen integration}%
\nomenclature[A]{$\psi(j)/i$}{Set/Index of sending ends of node $j$ in PDN}%
\nomenclature[A]{$\phi(j)/k$}{Set/Index of receiving ends of node $j$ in PDN}%
\nomenclature[A]{$\mathcal{T}/t$}{Set/Index of time periods}%
\nomenclature[A]{$\mathcal{U}$}{Uncertainty set}%
\nomenclature[B]{$M$}{Big-M factor}%
\nomenclature[B]{$\overline{G}_j$}{Number of MHERs allowed to connect to node $j$}%
\nomenclature[B]{$P^d_{j,t}/Q^d_{j,t}$}{Active/reactive power load of node $j$ at time $t$}%
\nomenclature[B]{$\overline{P}_m/\overline{Q}_m$}{Maximum active/reactive power output of MHER $m$}%
\nomenclature[B]{$\overline{P}_j^{sg}/\overline{Q}_j^{sg}$}{Maximum active/reactive power output of SG $j$}%
\nomenclature[B]{$R_{ij}/X_{ij}$}{Resistance/reactance of line $(i,j)$}%
\nomenclature[B]{$\overline{P}_{ij}/\overline{Q}_{ij}$}{Active/reactive power capacity of line $(i,j)$}%
\nomenclature[B]{$tr_{m,ij}$}{Travel time of MHER $m$ from node $i$ to $j$.}%
\nomenclature[B]{$\overline{H}_m$/$\underline{H}_m$}{Maximum/minimum hydrogen storage level of MHER $m$}%
\nomenclature[B]{$\overline{B}$}{Budget limit of proactive actions}%
\nomenclature[B]{$v_m$}{Fuel consumption rate of MHER $m$ when traveling}%
\nomenclature[B]{$\overline{V}_j/\underline{V}_j$}{Maximum/minimum voltage level of node $j$}%
\nomenclature[B]{$V_0$}{Reference voltage level of PND}%
\nomenclature[B]{$\hat{\Upsilon}$}{Target ratio of load restoration}%
\nomenclature[B]{$\eta_m$}{Generation efficiency of MHER $m$}%
\nomenclature[B]{$\xi$}{Hydrogen-to-power conversion factor}%
\nomenclature[B]{$\varpi_j$}{Priority weight of power demands at node $j$}%

\nomenclature[B]{$c_{ij}^H$}{Hardening cost factor of line $(i,j)$}%
\nomenclature[B]{$c_{m}^R$}{Renting cost factor of MHER $m$}%
\nomenclature[B]{$k$}{Max number of distribution lines that will be damaged}%
\nomenclature[C]{${u_{ij}}$}{Binary, 0 if line $(i,j)$ is damaged and 1 otherwise}%
\nomenclature[C]{$\chi_m$}{Binary, 1 if MHER $m$ is rented and 0 otherwise}%
\nomenclature[C]{$x_{ij}$}{Binary, 1 if line $(i,j)$ is hardened and 0 otherwise}%
\nomenclature[C]{$\gamma_{j,t}^m$}{Binary, 1 if MHER $m$ is connected to node $j$ at time $t$, 0 otherwise}%
\nomenclature[C]{$H_{m,t}$}{Hydrogen storage level of MHER $m$ at time $t$}%
\nomenclature[C]{${\omega_{ij,t}}$}{Binary, 1 if the line switch $(i,j)$ is closed at time $t$ and 0 otherwise}%
\nomenclature[C]{${P_{ij,t}}/{Q_{ij,t}}$}{Acitve/reactive power flow on line $(i,j)$ at time $t$}%
\nomenclature[C]{$pl_{j,t}/ql_{j,t}$}{Restored active/reactive loads in node $j$ at time $t$}%
\nomenclature[C]{$V_{j,t}$}{Voltage level in node $j$ at time $t$}%
\nomenclature[C]{$gp_{m,t}/gq_{m,t}$}{Acitve/reactive output of MHER $m$ at time $t$}%
\nomenclature[C]{$p_{j,t}^{sg}/q_{j,t}^{sg}$}{Acitve/reactive output of SG $j$ at time $t$}%
\nomenclature[C]{$f_{ij,t}$}{Fictitious flow on line $(i,j)$ at time $t$}%
\nomenclature[C]{$g_{j,t}$}{Ficticious power injection from source node $j$ at time $t$}%
\nomenclature[C]{${p_{j,t}}/{q_{j,t}}$}{Active/reactive power output of node $j$ at time $t$}%
\nomenclature[C]{${\varphi _{m,t}}$}{Binary, 1 if MHER $m$ is traveling at time $t$ and 0 otherwise}%
\nomenclature[C]{$H_{m,0}$}{Initial hydrogen storage level of MHER $m$}%

\section{Introduction}
\label{Sec:Intro}
\IEEEPARstart{T}{he} catastrophic events (e.g., hurricanes, wild fires and ice storms) may bring tremendous threats to the electrical power infrastructure, especially the power distribution network (PDN) \cite{poudyal2022risk}. For instance, the hurricane Ida and cold wave Uri  in 2021 rendered serious damages to the power distribution facilities in U.S., causing several billions of financial losses. To realize the resilient and reliable operation of PDNs, it demands for a flexible capability that facilitates disastrous prevention, resisting, adapting, and service restoration \cite{bie2017battling}. 

The resilience enhancement of PDNs can be achieved by reinforcing their physical structure or the adaptive scheduling of flexible resources under extreme disturbances, known as the ``infrastructural reinforcements'' and ``operational enhancements''  \cite{li2017networked}. The operational measures are contributed to alleviate the supply interruption incurred by power network outages, as well as the rapid load restoration \cite{li2017networked}. For the preventive scheduling of PDN before the disastrous events, a viable way is to deploy distributed energy resources (DERs), e.g., the micro turbines and fuel-cell stacks, and charge up the energy storage system \cite{wu2019microgrid,tobajas2022resilience}.
Also, the intentional microgrids islanding can be pre-planned to expand the preserving areas and power supply duration for critical loads \cite{jalilpoor2023resilient}. 
On the other hand, the post-event corrective measures can be conducted to facilitate the service restoration. By utilizing the smart measurement system and remotely-controlled switches (RCSs), the topology reconfiguration has become a common approach for a PDN to make fast emergency response and isolate the contingency areas \cite{ding2017resilient}. For further improving the efficiency of load restoration, the network reconfiguration is usually coordinated with the multi-period scheduling of DERs, which can form dynamic microgrids with flexible service boundaries \cite{lin2018tri}. Another way of leveraging the spatial flexibility is to deploy the mobile energy resources (MERs), conventionally driven by gasoline, diesel or battery energy storage, which can move among different contingency areas to sequentially perform on-emergency power supply. In contrast to stationary equipment, the MERs have a larger and changeable rescue coverage, so as to better handle the uncertainty of damaging spots \cite{kim2018enhancing,lei2018routing,wang2021scheduling,shi2024resilience}. Ref \cite{lei2018routing} focused on enhancing distribution system recovery through optimal routing and scheduling of MERs. Ref \cite{wang2021scheduling} introduced a concept of separable mobile energy storage systems, where the carrier and energy storage modules are treated as independent components, to further enhance the flexibility of MER applications. Additionally, Ref \cite{shi2024resilience} explored the use of MERs, via power-to-thermal conversion units, to support the recovery of multi-energy distribution systems. Recent years have witnessed the rapid proliferation of hydrogen energy technologies, which attracts an increasing attention for mobile hydrogen energy resources (MHERs) \cite{cao2022resilience,hassan2021hydrogen}. Compared to gasoline and diesel based MERs, the advantages of using MHERs including low carbon and pollutants emission, as well as a decreased noise level \cite{dong2023co}. Moreover, in contrast to mobile battery fleets with a high dependency on electricity networks, the MHERs are usually with more convenient and flexible refueling measures, including on-site water electrolysis, industrial hydrogen by-products, or delivery via existing natural gas pipelines \cite{kurtz2018renewable, liu2021resilient,jia2025decentralized}. Thus, the MHERs can provide more reliable and sustainable services under disastrous occasions. Also, in comparison to batteries, the hydrogen storage demonstrates several technical advantages, e.g., the higher storage content, shorter recharging time, and lower rates of degradation, which fits better for the long-duration power supply under extreme weather events \cite{cao2022resilience}. Actually, the industrial applications of MHERs have been reported in real disaster scenarios \cite{liang2018, office2019, foxnews2021}. For instance, during the strike of Typhoon ``Mangkhut'' in 2018, MHER ``Hornet'', equipped with a 120 kW hydrogen fuel cell system, successfully provided uninterrupted power supply to critical facilities for over six hours \cite{liang2018}. 	
	
Even though the collaboration of these operational measures would better help system recovery, they alone cannot induce the essential preparedness and structural resistance to disastrous events. In this regard, the infrastructural reinforcing measures, e.g., components hardening, vegetation management and changing network structure, are indispensable for the resilience enhancement of PDNs. The most common way to strengthen the power supply infrastructure is through network hardening, e.g., undergrounding power lines, poles/conductors fortifying, building redundant lines in vulnerable areas, adding windshield baffles \cite{yuan2016robust,Zhang2021TSG,Wang2019,9792201}. The optimal network hardening decision usually seeks for an improved load survivability subject to the cost budget restriction. The two-stage robust optimization (RO) method has been heavily adopted to address the uncertainty of network contingencies under worst-case consideration, e.g., in Refs \cite{yuan2016robust,Zhang2021TSG,tao2022multi,7514755,he2018robust}.   
Based on the $N-K$ contingency criterion, Ref \cite{yuan2016robust} proposed a two-stage RO model to coordinate the planning of line hardening and DER allocation with a multi-stage multi-zone uncertainty set to capture spatial and temporal dynamics of extreme weather events. On this basis, Ref \cite{Zhang2021TSG} further expanded the uncertainty set modeling considering the moving speed and angel of a typhoon. In addition to DERs, hardening is also coordinated with other measures to improve resilience. For instance, the decisions on network hardening and the deployment of RCSs are coordinated in \cite{9963579}. Ref \cite{tao2022multi} built a two-stage RO model to co-optimize line hardening, MERs scheduling and repair crews dispatching. Additionally, the Claude Shannon’s information theory was  adopted to formulate the uncertainty set for RO implementation in some studies \cite{7514755,he2018robust}. Recently, several studies have focused on the impact of proactive hardening decisions on the line fault probability. Ref \cite{zhang2022transmission} proposed a two-stage stochastic programming (SP) model to account for different line failures influenced by enhancement measures. Additionally, the hardening problem were also investigated in distributionally robust optimization (DRO) framework, aiming to identify the optimal hardening scheme against the worst-case distribution of contingencies \cite{li2023distributionally,zou2024two}.  

However, on the uncertainty modeling aspect, a fixed uncertainty set is generally defined based on prior knowledge in the traditional two-stage RO framework. But it is essentially static, which may not capture the system dynamics following the proactive actions for power resiliency. Indeed, adopting proper hardening measures would effectively decrease the damage risks of vulnerable network assets. So there actually exists a strong connection between hardening decisions and the concerned contingencies. Nevertheless, this connection is largely overlooked in current literature, where a static or decision-independent uncertainty (DIU) set is usually adopted. Clearly, the contingency of PDN with respect to hardening decisions is of an evident endogenous nature, also known as the \emph{decision-dependent uncertainty} (DDU) \cite{Zhang2022,zeng2022two,zhang2022transmission,li2023distributionally,zou2024two}. Even though the impacts of proactive decisions have been considered in a few existing formulations for distribution network hardening, they were solved heuristically \cite{7514755} or treated under DIU-based framework \cite{Zhang2021TSG}. There still lacks systematic study and deep insights for the endogenous uncertainty representation in the context of an RO-based hardening problem. Actually, without properly characterizing the structural properties of a DDU set, it may lose the robust feasibility and lead to overly conservative results. On the other hand, we consider two sequentially correlated stages for the resilient operation of PDN, i.e., the pre-event preparation and post-event service restoration. At the preventive stage, the structural reinforcement (e.g., network hardening and DERs investment) is a fundamental way to improve system's preparedness to the disastrous events. But the proactive measures are usually restricted by a limited cost budget, which can hardly cover all those vulnerable regions. So, it is necessary to take operational remedy measures (e.g., the on-emergency scheduling of DGs and energy storage devices, as well as microgrids formation) at the post-event stage. Particularly, the dynamic re-routing and energy scheduling of mobile energy resources could bring additional spatial flexibility through reliable roadway transportation \cite{lei2018routing,cao2022resilience,wang2021scheduling}, which may significantly enlarge the coverage of resilience measures. Hence,  the coordination of long-term network hardening and short-term mobile resources deployment provides a rather cost-effective way to improve the power grids preparedness and responsiveness under contingency occasions.
	
Based on the above analysis, this paper presents a co-planning framework of network hardening and mobile energy resource scheduling for seeking a cost-effective way to facilitate the resilient operations of PDN. The on-emergency routing and energy scheduling of MHERs are considered, as a zero-carbon alternative for conventional MERs (e.g., powered by gasoline, diesels and batteries). Also, we consider two types of uncertain factors that may influence the load survivability level under catastrophic occasions. The former one is the discrete uncertainty of distribution lines, which can be significantly influenced by proactive hardening actions. The latter one is the continuous randomness of available storage levels of MHERs, which depends on rental decisions and could affect the effectiveness of disaster-relief strategies. To explicitly capture these endogenous uncertainties, a closed-form decision-dependent uncertainty (DDU) set is developed, which has a mixed-integer structure that is very new in the literature of power system research. Then, we propose a two-stage DDU-based RO formulation with $\min-\max$ resilience constraint and mixed-integer programming (MIP) recourse problem for enhancing system's preparedness and resistance against disastrous events. The first-stage problem is to proactively optimize the hardening scheme of PDN, along with the pre-allocation plan of MHERs. In the second-stage, the bi-level constraint is incorporated to guarantee a satisfactory load survivability level under the DDU-based worst-case considerations. Besides, the multi-period routing of MHERs and network reconfiguration are co-scheduled to further reduce the resilience cost. 

In contrast to the classical form of RO, the proposed model has three computational challenges. First, the shape of uncertainty set is dynamically changed by the proactive hardening and MHERs' pre-allocation decisions. Second, the mixed-integer structure significantly increases the computational difficulty. Third, although a few recent approaches are designed to address the DDU-based RO problems, e.g., the modified Benders decomposition (e.g., Refs \cite{zhang2021robust,Zhang2022}), dual C\&CG \cite{tan2024robust} and parametric C\&CG \cite{zeng2022two} algorithms, they all rely on the acquisition of exact dual variables from the recourse problem, i.e., strong duality must hold for the innermost problem. However, for our specific two-stage formulation, the dual information cannot be precisely extracted because of the mixed-integer recourse structure. As pointed out in Ref \cite{tan2024robust}, there is no existing algorithm that can handle the DDU-based RO with mixed-integer recourse problems. Another computational challenge is induced by the $\min-\max$ resilience constraint, so we have to exactly solve those bi-level MIP subproblems rather than using the approximation strategies based on LP relaxation \cite{zeng2022two}. To address the aforementioned computational challenges, we develop a nested solution scheme with strong enhancements based on the algorithm logic of parametric C\&CG \cite{zeng2022two,wang2023computing}, which offers a different perspective on \cite{tan2024robust}' statement. 
By leveraging the structural property of the DDU set as well as the combination of discrete recourse decisions and the corresponding extreme points, the nontrivial scenarios in the dynamical DDU set can be precisely characterized and the nested solution algorithm can be proved to have finite convergence.  Additionally, nontrivial enhancement strategies, i.e., the inclusion of $\mathcal{OU}'(\bm{x})$ (an additional optimal solution set), and the warm-start of inner-loop C\&CG, are designed and implemented to improve the convergence performance. Hence, it provides a strong and exact solution tool to handle this challenging problem.

In comparison to the existing studies, the major contributions of this paper are listed as below:
\begin{enumerate}
	\item  A mixed-integer DDU set is developed to reveal two pairs of practical dependence, i.e., between proactive network hardening and contingency uncertainties,  as well as the MHERs' pre-allocation and their available storage levels. 
	\item A DDU-based robust MIP model is proposed for the resilient co-optimization of distribution network hardening and MHERs' scheduling. By imposing the $\min-\max$ resilience constraint, the multi-period routing of MHERs and network reconfiguration are synergized in the innermost problem to achieve a guaranteed load survivability level under the worst-case scenario.
	\item To address the computational challenges of DDU-based RO with MIP recourse problem, a nested parametric C\&CG (N-PC\&CG) algorithm is designed and improved by two enhancement strategies. By making use of the structure of DDU set as well as the implicit enumeration of discrete recourse variables and corresponding extreme points, the proposed model can be exactly and effectively solved.  
\end{enumerate}

The rest part of this paper is organized as below. Section II presents the general form of DDU-based and resilience-constrained RO model. Section III gives its detailed mathematical representation. Section IV develops the N-PC\&CG  algorithm with an enhanced customization. Section V shows the results of numerical studies. Finally, conclusions are drawn in Section VI. 

\section{Resilience-Constrained DDU-RO Model}
\label{Sec:Model}
We consider a DDU-based two-stage RO framework for enhancing the resilience of PDN. The first-stage problem is to minimize the accumulated cost of proactive distribution lines hardening (denoted by a binary vector $\bm{x}$) and mobile resource deployment (represented by a binary vector $\bm{z}$ indicating the renting and preventive allocation status of MHERs), which are determined before the realization of any network contingency (represented by a binary vector $\bm{u}$). 
For the second-stage problem, the post-event on-emergency decisions of multi-period MHERs' re-routing (denoted by a binary vector $\bm{\gamma}$), network reconfiguration (denoted by a binary vector $\bm{\omega}$) as well as the coordinated system operation (represented by a vector of continuous variables $\bm{y}$ indicating the sequential energy scheduling of MHERs and power flow status of reconfigurable PDN) are optimized through a bi-level MIP formulation. Particularly, a $\min-\max$ resilience constraint is enforced to fulfill the load survivability requirement (i.e., no less than a target level $\hat{\Upsilon}$) under the worst-case contingency scenario. 

Let $\mathcal{U}(\bm{x},\bm{z})$ denote the decision-dependent contingency set, a point to set map for $\bm{u}$ and $\bm{H}_0$. We present the compact form of our resilience-based co-planning model as below: 
\setlength{\arraycolsep}{-0.3em}
\begin{eqnarray}
	&&{\bf DDU-RRO}: \ \Gamma=\min_{\bm{x},\bm{z}\in \mathcal{X}}  \ \bm{c}^T\bm{x}+ \bm{d}^T\bm{z} \label{DDU-RRO1}\\
	&&\quad\quad\quad\quad s.t.,\min_{\bm{u}, \bm{H}_0\in \mathcal{U}(\bm{x},\bm{z})} \max_{\bm{\gamma},\bm{\omega},\bm{y}\in \mathcal{Q}(\bm{z},\bm{u},\bm{H}_0) } \bm{q}^T\bm{y} \geq \hat{\Upsilon} \label{DDU-RRO2}
\end{eqnarray}
where the planning objective in \eqref{DDU-RRO1} is to minimize the system reinforcement cost (SRC) for the upcoming disastrous events. Also, the resilience constraint in \eqref{DDU-RRO2} is constructed based on an affine function $\bm{q}^T\bm{y}$ that evaluates the loads supplying level under critical network damages. 

We next define the constraints set concerning the proactive and post-event corrective actions. The first-stage constraints in $\mathcal{X}$ capture the long-term and strategic restrictions, e.g., the geographical and cost budget constraint, for network hardening and mobile resources' pre-allocation plans. Note that $\bm{A}$, $\bm{D}$ and $\bm{b}$ represent the coefficient matrices and right-hand side parameter vector in the first-stage constraints.
\setlength{\arraycolsep}{-0.3em}
\begin{equation}
	\label{FS} 
	\mathcal{X} = \left\lbrace (\bm{x},\bm{z}) \in \{0,1\}^{n_1} \times \{0,1\}^{n_2}: \ \bm{A}\bm{x} +\bm{D}\bm{z} \leq \bm{b} \right\rbrace 
\end{equation}

Note that the connectivity of distribution lines under disastrous occasions can be endogenously affected by the hardening decision. For instance, burying the overhead lines underground or reinforcing building material can largely protect them from the extreme weather events. Thus a rationale assumption can be made, under these specific real-world instances, that the distribution lines is considered to be outage-free once they are hardened. On the other hand, even though the hydrogen refueling process could be fast, the remaining hydrogen storage levels of MHERs may be random due to two practical reasons: 1) the limited deployment of hydrogen refueling infrastructure, driven by high capital costs \cite{greene2020challenges} may result in MHERs being far away from available refueling sites when disasters occur. 2) the road disruptions caused by extreme events (e.g., strong winds and floods) can prolong transportation time \cite{cao2022resilience}, leading to uncertain hydrogen consumption. Considering these two types of uncertain factors, we have the DDU set as:
\setlength{\arraycolsep}{-0.3em}
\begin{eqnarray}
	&&\mathcal{U}(\bm{x},\bm{z})=\{ (\bm{u},\bm{H}_0)\in \{0,1\}^{m}\times \mathcal{R}^{a} : \nonumber \\ 
	&&\quad \quad \quad \quad \quad \quad \quad  \bm{1}^T \bm{u} \geq N-k \label{DDU1}  \\
	&&\quad \quad \quad \quad \quad \quad \quad  \bm{u}\geq \bm{x}\label{DDU2} \\
	&&\quad \quad \quad \quad \quad \quad \quad  \sigma_1 \bm{z} \circ \overline{\bm{H}} \leq \bm{H}_0 \leq \bm{z} \circ \overline{\bm{H}}  \label{DDU3} \\
	&&\quad \quad \quad \quad \quad \quad \quad  \bm{1}^T \bm{H}_0 \geq \sigma_2 {\bm{z}}^T  \overline{\bm{H}}  \} \label{DDU4}
\end{eqnarray}
where the $N-k$ criteria is applied in \eqref{DDU1} to restrict the number of damaged feeders \cite{Wang2019}. Let $x_{ij}$ and $u_{ij}$ be the $(i,j)-$th component of $\bm{x}$ and $\bm{u}$ respectively. $u_{ij}=0$ indicates that line $(i,j)$ is damaged, and $u_{ij}=1$ otherwise. Also, the impact of hardening decisions is imposed by \eqref{DDU2}. As long as $x_{ij}=1$, line $(i,j)$ is under protection, so that $u_{ij}$ will be fixed at 1. In other instances, they are free uncertain variables. So it well captures the bilateral dependence between hardening decisions and contingency uncertainties. Additionally, the dependence relations between the initial hydrogen storage level of MHERs ($\bm{H}_0$) and their renting decisions $\bm{z}$ are described in \eqref{DDU3} and \eqref{DDU4}. Note that $\bm{a}\circ \bm{b}$ represents a Hadamard product of vectors $\bm{a}$ and $\bm{b}$. We assume that the initial hydrogen storage level of MHERs is random within a specific range, as in \eqref{DDU3}, where parameters $\overline{\bm{H}}$ and $\sigma_1$ represent the full hydrogen storage levels of MHERs and their initial lower bound. Moreover, for some MHERs that charge more for on-emergency reserve services (i.e., a higher renting cost), the rental provider may guarantee that the total hydrogen storage of these MHERs will be higher than a required level (i.e., $\sigma_2$), which is imposed by \eqref{DDU4}.

\begin{remark}
	\label{rmk:1}
	In real situations, the aging status and geographical locations of different components in PDN differ, as does the extent to which they are influenced by disasters. So, assuming that all lines to be equally damaged would lead to an overly conservative hardening plan. Suppose that the fault probability of each line can be obtained through proper fragility modeling \cite{7514755}. Then, we can introduce a threshold to determine which lines could be damaged, thereby reducing the conservatism of uncertainty model \cite{yang2024resilient}. In this paper, we consider that \eqref{DDU1}-\eqref{DDU2} are defined on a set of vulnerable lines, while the lines that are invulnerable would not be damaged.
\end{remark}

Additionally, the feasible set of corrective variables for the second-stage recourse decisions is presented as below:
\begin{eqnarray}
	&&\mathcal{Q}(\bm{z},\bm{u},\bm{H}_0)=\{(\bm{\gamma}, \bm{\omega},\bm{y})\in \{0,1\}^{r_1}\times \{0,1\}^{r_2}\times \mathbb{R}^p: \label{ST1}\\
	&&\quad \quad \quad \quad \quad \bm{W}\bm{\gamma}+\bm{F}\bm{y}\geq \bm{f}-\bm{G}\bm{H}_0 \label{ST2}\\
	&&\quad \quad \quad \quad \quad \bm{H}\bm{y}+\bm{L}\bm{\omega}\bm{u}\geq \bm{h}\} \label{ST3} \\
	&&\quad \quad \quad \quad \quad \bm{J}_1\bm{\gamma}+\bm{J}_2\bm{\omega}\geq \bm{l}-\bm{T}\bm{z} \}  \label{ST4}
\end{eqnarray}
where $\bm{W}$, $\bm{F}$, $\bm{G}$, $\bm{H}$, $\bm{L}$, $\bm{J}_1$, $\bm{J}_2$, $\bm{T}$ are the matrix coefficients in the second-stage constraints, and $\bm{f}$, $\bm{h}$, $\bm{l}$ are the corresponding right-hand side parameter vectors. The multi-period routing of MHERs, network reconfiguration and system operation are denoted by variables $\bm{\gamma}$, $\bm{\omega}$ and $\bm{y}$, respectively. Specifically, Eq. \eqref{ST2} associates the resource relocation and energy scheduling with the available storage levels, corresponding to \eqref{2nd:CONS5}-\eqref{2nd:CONS13} and  \eqref{2nd:CONS21}-\eqref{2nd:CONS19-1}. The consecutive power flow states of PDN at post-event stage are constrained by \eqref{ST3}, corresponding to \eqref{2nd:CONS14}-\eqref{2nd:CONS26} and \eqref{2nd:CONS9}-\eqref{2nd:CONS11}. The constraints that only contain binary variables (i.e., $\bm{\gamma}$ and $\bm{\omega}$) are denoted by \eqref{ST4}, corresponding to \eqref{2nd:CONS1}-\eqref{2nd:CONS4} and \eqref{2nd:CONS8}. Notice again that the network integrity of the hardened parts can be preserved, as in \eqref{DDU1}-\eqref{DDU2}. So the influence of network hardening is implicitly considered in \eqref{ST3} with incorporating $\bm{u}$ through the DDU set. On the other hand, the damaged network parts under extreme disturbances can be isolated through topology reconfiguration. The formation of these dynamic microgrids in coordination with the flexible multi-area relocation of MHERs can decrease the risk cost of sustaining critical loads under emergencies. 

\begin{remark}
	\label{rmk:2}
	Different from the classical RO with a fixed uncertainty set, which overlooks the impact of proactive decisions on uncertainty realization, the DDU-based two-stage RO model in \eqref{DDU-RRO1}-\eqref{DDU-RRO2} captures the endogenous effects of network hardening and mobile resources allocation. Moreover, our problem holds a special recourse structure, i.e., with $\min-\max$ bi-level constraint. It helps achieve a required resilience level under DDU-based worst-case considerations. The detailed expressions for \eqref{DDU-RRO1}-\eqref{ST4} will be presented in the next section.
\end{remark} 

\section{Detailed Mathematical Formulation}
\label{Sec:Details}
The compact matrix formulation presented in Section II provides a high-level abstraction of the optimization model, revealing the complex interdependencies between decision variables and constraints. In the following, we elaborate on the detailed mathematical expressions of \eqref{DDU-RRO1}-\eqref{ST4}, organized according to their roles and functionalities with physical interpretations.

\subsection{Objective Function: Minimization of Reinforcement Cost}
The planning objective of minimizing SRC corresponding to \eqref{DDU-RRO1} can be expressed as:
\begin{eqnarray}
	&&\mathop{\min}_{\bm{x},\bm{z} \in \mathcal{X}} \ \sum_{(i,j)\in \mathcal{L}} c^{H}_{ij} x_{ij}  + \sum_{m \in \mathcal{M}} c_m^R \chi_m \label{OBJ} \\
	&&\bm{x}=\left\lbrace x_{ij} | \forall (i,j) \in \mathcal{L} \right\rbrace \subseteq \{0,1\}^{n_1} \label{var1} \\
	&&\bm{z}=\left\lbrace \chi_m, \gamma _{j,0}^{m} | \forall m \in \mathcal{M}, \forall j \in \mathcal{N}^{eh}\right\rbrace \subseteq \{0,1\}^{n_2}   \label{var2}  
\end{eqnarray}

Note that the first term in \eqref{OBJ} represents the expenditures for network hardening, e.g., the construction cost of reinforcing the poles of distribution lines \cite{Zhang2021TSG}.
The second term is the cost of renting MHERs for emergency use. The cost coefficient $c_m^R$ can be determined according to both the renting duration and hydrogen fuels consumption. \vspace{-5pt}

\subsection{Constraints of Proactive Actions}
The proactive decisions on network hardening ($x_{ij}$) as well as the renting and pre-scheduling of MHERs ($\chi_m$ and $\gamma _{j,0}^{m}$) are constrained as in set $\mathcal{X}$, corresponding to \eqref{FS}, which includes 
\setlength{\arraycolsep}{-0.2em}
\begin{eqnarray}
	&&\sum_{(i,j)\in \mathcal{L}} c^{H}_{ij} x_{ij}  + \sum_{m \in \mathcal{M}} c_m^R \chi_m \leq \overline{B} \label{lst:budget} \\
	&&\sum\limits_{j\in {{\mathsf{\mathcal{N}}}^{eh}}}{\gamma _{j,0}^{m}}\leq {{\chi }_{m}},\quad \forall m\in \mathsf{\mathcal{M}} \label{1st:CONS2} \\
	&&\sum\limits_{m\in \mathsf{\mathcal{M}}}{\gamma _{i,0}^{m}}\leq \overline{G}_j,\quad \forall j\in {{\mathsf{\mathcal{N}}}^{eh}} \label{1st:CONS3}
\end{eqnarray}
where the total reinforcement cost is constrained by a budget cap in \eqref{lst:budget}. Also, the renting and pre-allocation decisions of MHERs are constrained by \eqref{1st:CONS2}-\eqref{1st:CONS3}. Each MHER, if rented, can be connected to at most one EH node (with pre-installed vehicle-to-grid (V2G) interfaces), as in \eqref{1st:CONS2}. For each EH node, the number of deployed MHERs could be restricted by parking space and the capacity of V2G facilities, as in \eqref{1st:CONS3}. 
\vspace{-10pt}

\subsection{Constraints of MHERs' Re-routing and On-Emergency Energy Scheduling}
At the post-event stage, those pre-allocated MHERs would be re-routed among different EH nodes for service restoration of outage areas. The multi-period routing and scheduling constraints of MHERs in \eqref{ST2} and \eqref{ST4} can be elaborated as:
\begin{eqnarray}
	&&\sum_{j\in \mathcal{N}^{eh}} \gamma_{j,t}^m \leq \chi_m, \quad \forall m \in \mathcal{M}, \forall t  \in \mathcal{T} \label{2nd:CONS1} \\
	&&\sum_{m\in \mathcal{M}} \gamma_{j,t}^m \leq \overline{G}_j, \quad \forall j \in \mathcal{N}^{eh}, \forall t  \in \mathcal{T} \label{2nd:CONS2} \\
	&&\varphi_{m,t}=\chi_m - \sum_{j\in \mathcal{N}^{eh}} \gamma_{j,t}^m, \quad \forall m \in \mathcal{M}, \forall t  \in \mathcal{T} \label{2nd:CONS3} \\
	&&\begin{aligned}
		\gamma_{i,t}^m + \gamma_{j,t+\tau}^m \leq \chi_m, \quad \forall m \in \mathcal{M}, \forall i,j \in \mathcal{N}^{eh}, \\  \forall \tau \leq tr_{m,ij},  \forall t+\tau \leq \left| \mathcal{T}\right| \label{2nd:CONS4}
	\end{aligned}\\
	&&{H}_{m,t}={H}_{m,t-1}-gp_{m,t}/(\eta_{m}\xi)-v_m\varphi_{m,t}, \forall m\in \mathcal{M},\forall t \label{2nd:CONS5}\\
	&&\underline{H}_m \leq {H}_{m,t}\leq \overline{H}_m, \quad \forall m\in \mathcal{M}, \forall t\in \mathsf{\mathcal{T}} \label{2nd:CONS6} \\
	&&0\le gp_{m,t}\le \overline{P}_m,\quad\forall m\in \mathsf{\mathcal{M}},\forall t\in \mathsf{\mathcal{T}}\label{2nd:CONS7-1} \\
	&&0\le gq_{m,t}\le \overline{Q}_m,\quad\forall m\in \mathsf{\mathcal{M}},\forall t\in \mathsf{\mathcal{T}}\label{2nd:CONS7-2}	
\end{eqnarray}

The requirements for MHERs' re-routing are imposed as in \eqref{2nd:CONS1}-\eqref{2nd:CONS4}. The meanings of constraints \eqref{2nd:CONS1}-\eqref{2nd:CONS2} are analogous to those of \eqref{1st:CONS2}-\eqref{1st:CONS3}. Note that for an activated MHER, the states of grid integration and roadway traveling are mutually exclusive in each time period. So, it must be traveling if not parked with any EH node, as in \eqref{2nd:CONS3}. Additionally, the movement of MHERs among different EH nodes must take necessary transportation time, which is imposed in \eqref{2nd:CONS4} \cite{lei2018routing}. Also, the hydrogen storage level of each fleet will decrease once it travels between two nodes or discharges via the V2G interface, as defined in  \eqref{2nd:CONS5}. Eq. \eqref{2nd:CONS6} restricts the ranges of remaining hydrogen storage levels of MHERs. The active/reactive power outputs of MHERs are restricted by their rated power, as expressed in \eqref{2nd:CONS7-1} and \eqref{2nd:CONS7-2} \cite{cao2022resilience}. 
Note that the initial storage levels of MHERs (${H}_{m,0}$ for all $m\in \mathsf{\mathcal{M}}$) are uncertain variables, which are defined as in  \eqref{DDU3} and \eqref{DDU4}.

\subsection{Constraints of Reconfigurable Distribution Network}
The post-event network operation with adaptive topology control in \eqref{ST2}-\eqref{ST3} can be elaborated as in the following: 
\setlength{\arraycolsep}{-0.4em}
\begin{eqnarray}
	&& \sum\limits_{i \in \psi (j)} {{P_{ij,t}} - } \sum\limits_{k \in \phi (j)} {{P_{jk,t}}}  = pl_{j,,t}-p_{j,t}, \quad \forall j \in {\cal N},\forall t \in \mathcal{T} \label{2nd:CONS12} \\
	&& \sum\limits_{i \in \psi (j)} {{Q_{ij,t}} - } \sum\limits_{k \in \phi (j)} {{Q_{jk,t}}}  = ql_{j,t}-	q_{j,t}, \quad \forall j \in {\cal N},\forall t \in \mathcal{T} \label{2nd:CONS13} \\
	&&V_{i,t}-V_{j,t} \le \left(r_{ij}P_{ij,t}+x_{ij}Q_{ij,t}\right) /V_0+M(1-u_{ij}\omega_{ij,t}), \nonumber \\
	&&\quad \quad \quad \quad \quad \quad \quad \quad \quad \quad \quad \quad \quad \quad \forall (i,j) \in \mathcal{L}, \forall t \in \mathcal{T}  \label{2nd:CONS14} \\
	&&V_{i,t}-V_{j,t} \ge \left(R_{ij}P_{ij,t}+X_{ij}Q_{ij,t}\right) /V_0+M(u_{ij}\omega_{ij,t}-1), \nonumber \\
	&&\quad \quad \quad \quad \quad \quad \quad \quad \quad \quad \quad \quad \quad \quad \forall (i,j) \in \mathcal{L}, \forall t \in \mathcal{T}  \label{2nd:CONS15}\\
	&&-\overline{P}_{ij} u_{ij} \omega_{ij,t} \leq P_{ij,t}\leq \overline{P}_{ij} u_{ij} \omega_{ij,t},  \forall (i,j)\in \mathcal{L}, \forall t \in \mathcal{T} \label{2nd:CONS24}\\
	&&-\overline{Q}_{ij} u_{ij} \omega_{ij,t} \leq Q_{ij,t}\leq \overline{Q}_{ij} u_{ij} \omega_{ij,t},  \forall (i,j)\in \mathcal{L}, \forall t \in \mathcal{T} \label{2nd:CONS25}\\
	&&\underline{V}_{j}\leq V_{j,t}\leq \overline{V}_{j}, \quad \forall j\in\mathcal{N}, \forall t \in \mathcal{T} \label{2nd:CONS26} \\
	&&\sum\limits_{(i,j) \in \mathsf{\mathcal{L}}}{{\omega_{ij,t}}=\left| \mathcal{N}\right|-\left|\mathcal{N}^{sub}\right|,\quad \forall t\in \mathsf{\mathcal{T}}} \label{2nd:CONS8}\\
	&&\sum\limits_{i\in \psi(j)}f_{ij,t}-\sum\limits_{k\in \phi(j)}f_{jk,t}=1, \forall t\in \mathsf{\mathcal{T}},\forall j\in \mathcal{N} \backslash \{\mathcal{N}^{sub},\mathcal{N}^{eh}\} \label{2nd:CONS9}\\
	&&\sum\limits_{i\in \psi(j)}f_{ij,t}-\sum\limits_{k\in \phi(j)}f_{jk,t}=-g_j, \forall t\in \mathsf{\mathcal{T}},\forall j\in \mathsf{\mathcal{N}}^{eh}\cup \mathsf{\mathcal{N}}^{sub} \label{2nd:CONS10}\\
	&&-M_1{\omega}_{ij,t}\le {{f}_{ij,t}}\le M_1{\omega}_{ij,t},\quad \forall t\in \mathsf{\mathcal{T}},\forall (i,j)\in \mathsf{\mathcal{L}} \label{2nd:CONS11}
\end{eqnarray}

The linear \emph{DistFlow} model \cite{baran1989reconfig} is leveraged and modified to characterize the operations of reconfigurable distribution network. The nodal active/reactive power balance constraints are imposed through \eqref{2nd:CONS12} and \eqref{2nd:CONS13}. The physical relationship of voltage levels with the line flow status is defined as in \eqref{2nd:CONS14}-\eqref{2nd:CONS15}. A very large positive constant $M$ is introduced to relax the constraints associated with disconnected lines.
Also, for ensuring the operational security, the active and reactive power flows on each line are constrained by \eqref{2nd:CONS24} and \eqref{2nd:CONS25}. The nodal voltage level is constrained by \eqref{2nd:CONS26}.

The enforcement of constraints \eqref{2nd:CONS14}-\eqref{2nd:CONS25} are correlated with the real-time topology of PDN, which can be controlled by dynamically changing the on/off status of smart line switches (indicated by binary variables ${\omega}_{ij,t}$) based on the post-event network integrity (represented by binary variables $u_{ij}$) \cite{baran1989reconfig}. When $u_{ij}=0$ or $\omega_{ij}=0$, the power flows on line $(i,j)$ will be forced as zero. Constraints \eqref{2nd:CONS8}-\eqref{2nd:CONS11} are included to maintain the radiality and connectivity of reconfigured network. Specifically, the radiality condition is placed by \eqref{2nd:CONS8}. The formulation of fictitious power flows is introduced in \eqref{2nd:CONS8}-\eqref{2nd:CONS11} to guarantee that the load nodes in a reconfigured topology can be connective to at least one source node (i.e., the node with DER or distribution substation). Detailed explanation can be found in Refs. \cite{cao2022resilience,Cao2019Reorganization}. 

Furthermore, for the sake of a stable and positive recovery process, the rate of load restoration should be non-decreasing with the time evolving, as in \eqref{2nd:CONS21}-\eqref{2nd:CONS23}.
\setlength{\arraycolsep}{-0.2em}
\begin{eqnarray}
	&&0 \le pl_{j,t} \le P_{j,t}^d, \forall j\in \mathsf{\mathcal{N}},\forall t \in \mathcal{T} \label{2nd:CONS21} \\
	&&pl_{j,t-1}/P_{j,t-1}^d\le pl_{j,t}/P_{j,t}^d,\forall j\in \mathsf{\mathcal{N}},\forall t \in \mathcal{T} \label{2nd:CONS22} \\
	&&ql_{j,t}=(Q_{j,t}^d/P_{j,t}^d)\cdot pl_{j,t},\forall i\in \mathsf{\mathcal{N}},\forall t\in \mathsf{\mathcal{T}} \label{2nd:CONS23} 
\end{eqnarray} 
where the active and reactive power loads ($pl_{j,t}$/$ql_{j,t}$) can be satisfied by both the nodal generation and the power injection from the utility grid (i.e., the power import through distribution substations). The constraints for the power outputs of different node types can be written as:
\setlength{\arraycolsep}{-0.2em}
\begin{eqnarray}
	&&0 \le p_{j,t} \le \overline{P}^{sub}, \quad \forall j \in \mathcal{N}^{sub},  \forall t \in \mathcal{T}  \label{2nd:CONS16}\\
	&&0 \le q_{j,t} \le \overline{Q}^{sub}, \quad \forall j \in \mathcal{N}^{sub},  \forall t \in \mathcal{T}  \label{2nd:CONS17}\\
	&&p_{j,t}=p^{sg}_{j,t}+\sum_{m\in \mathcal{M}} \gamma^m_{j,t} gp_{m,t}, \quad \forall j\in \mathcal{N}^{eh}, \forall t \in \mathcal{T} \label{2nd:CONS18}\\
	&&q_{j,t}=q^{sg}_{j,t}+\sum_{m\in \mathcal{M}} \gamma^m_{j,t} gq_{m,t}, \quad \forall j\in \mathcal{N}^{eh}, \forall t \in \mathcal{T} \label{2nd:CONS19} \\
	&&0\leq p^{sg}_{j,t}\leq \overline{P}^{sg}_j,\quad \forall j\in \mathcal{N}^{eh}, \forall t \in \mathcal{T}\label{2nd:CONS18-1}\\
	&&0\leq q^{sg}_{j,t}\leq \overline{Q}^{sg}_j,\quad \forall j\in \mathcal{N}^{eh}, \forall t \in \mathcal{T}\label{2nd:CONS19-1}
\end{eqnarray}
where \eqref{2nd:CONS16}-\eqref{2nd:CONS17} are importing power constraints for the substation nodes. For those EH nodes, the self-generation capability for emergency use can be achieved through stationary distributed generators (e.g., FCs and micro turbines) and the integration of MHERs, as defined in \eqref{2nd:CONS18} and \eqref{2nd:CONS19}. Besides, the active and reactive power outputs of stationary generators (SGs) are constrained by \eqref{2nd:CONS18-1}-\eqref{2nd:CONS19-1}. 

\begin{remark}
	\label{rmk:3}
    Note that \eqref{2nd:CONS18} and \eqref{2nd:CONS19} introduce bilinear terms, 
    which could make our problem intractable for large-scale problems. Following the idea of ``big-M'' method, these bilinear parts can be linearized by adding new variables and a couple of auxiliary constraints. For instance, we consider $\gamma^m_{j,t} gp_{m,t}(=GP_{j,t}^{m})$ in \eqref{2nd:CONS18}, the linearization scheme can be demonstrated as below:
	\begin{eqnarray}
	&&0\le GP_{j,t}^{m}\le \gamma _{j,t}^{m}\overline{P}_m \label{2nd:CONS27}\\
	&&gp_{m,t}+(\gamma _{j,t}^{m}-1)\overline{P}_m \le GP_{j,t}^{m}\le gp_{m,t} \label{2nd:CONS28}	
	\end{eqnarray}
    where $GP_{j,t}^{m}$ becomes $0$ once $\gamma _{j,t}^{m}=0$ due to \eqref{2nd:CONS27}. Together with \eqref{2nd:CONS28}, we have $GP_{j,t}^{m}=gp_{m,t}$ when $\gamma _{j,t}^{m}=1$, which is equivalent to the original formulation.
\end{remark} \vspace{-10pt}

\subsection{Resilience Constraint}
Finally, the $\min-\max$ resilience constraint in \eqref{DDU-RRO2} can be elaborated as follows:
\begin{eqnarray}
	&&\min_{\bm{u},\bm{H}_0\in \mathcal{U}(\bm{x},\bm{z})} \max_{\bm{\gamma},\bm{\omega},\bm{y}\in \mathcal{Q}(\bm{z},\bm{u},\bm{H}_0) } \frac {\sum_{t\in \mathcal{T}} \sum_{j\in {\cal N}} \varpi_j pl_{j,t}}{\sum_{t\in \mathcal{T}} \sum_{j\in {{\cal N}}} \varpi_j P^d_{j,t}}\geq \hat{\Upsilon} \label{2nd:RC}\\
	&&\bm{\gamma}=\left\lbrace \gamma_{j,t}^m|\forall j \in \mathcal{N}^{eh}, \forall m \in \mathcal{M}, \forall t \in \mathcal{T} \right\rbrace \subseteq \{0,1\}^{r_1}    \label{2nd:VAR1}\\
	&&\bm{\omega}=\left\lbrace \omega_{ij,t}|\forall (i,j)\in \mathcal{L},\forall t \in \mathcal{T} \right\rbrace \subseteq \{0,1\}^{r_2}    \label{2nd:VAR2}\\
	&&\bm{y}=\left\lbrace P_{ij,t},Q_{ij,t},V_{j,t}, pl_{j,t},ql_{j,t},p_{j,t},q_{j,t}|\forall (i,j)\in \mathcal{L},\forall t\in \mathcal{T}\right\rbrace \nonumber\\
	&&\quad\ \bigcup \left\lbrace gp_{m,t},gq_{m,t},H_{m,t}|\forall m\in \mathcal{M},\forall t\in \mathcal{T}\right\rbrace \subseteq \mathbb{R}^p  \label{2nd:VAR3}
\end{eqnarray}
where $\varpi_j$ is the priority weight for restoring the power loads at node $j$. As in \eqref{2nd:RC}, the service restoration ratio is expressed as the percentage of total restored loads over the rescuing periods to the weighted summation of nominal load demands, which should be no lower than a pre-specified target under the DDU-based worst-case scenario. Besides, the detailed variables in recourse decision vectors $\bm{\gamma}$, $\bm{\omega}$ and $\bm{y}$ are listed in \eqref{2nd:VAR1}-\eqref{2nd:VAR3}. 

\section{Solution Algorithm}
\label{Sec:Algorithm}
Note that ${\bf DDU-RRO}$ in \eqref{DDU-RRO1}-\eqref{ST3} is a very challenging RO formulation integrating $\min-\max$ constraint, mixed-integer recourse as well as the endogenous uncertainty representation. As the DDU set $\mathcal{U}(\bm{x},\bm{z})$ depends on $\bm{x}$ and $\bm{z}$, its shape varies with the first-stage decisions dynamically. The extreme point identified as the ``worst-case'' scenario in $\mathcal{U}(\bm{x}^1,\bm{z}^1)$ will generally not be an extreme point of $\mathcal{U}(\bm{x}^2,\bm{z}^2)$ when $(\bm{x}^1,\bm{z}^1) \neq (\bm{x}^2,\bm{z}^2)$. In fact, it may not even belong to $\mathcal{U}(\bm{x}^2,\bm{z}^2)$. The changeable characteristic of DDU set makes the classical nested C\&CG method \cite{zhao2012exact} inapplicable. Furthermore, existing algorithms (e.g., Refs \cite{zhang2021robust,zeng2022two,Zhang2022,tan2024robust}) designed for DDU-based RO problems are also unable to address the proposed model, primarily because the dual information associated with the innermost problem cannot be precisely derived under the MIP recourse structure. To efficiently solve the proposed model, we develop a nested customization of parametric C\&CG (PC\&CG). By enumerating the combinations of discrete recourse variables and corresponding extreme points, the proposed model can be exactly solved.	

\subsection{$\mathcal{OU}$ Formulation for N-PC\&CG}
\label{sec:4A}
For a given first-stage solution $\left({\bm{x}},{\bm{z}}\right)$, we first consider a $\min-\max$ subproblem by extracting the left-hand-side bi-level optimization problem of \eqref{DDU-RRO2} and its constraints presented in \eqref{ST1}-\eqref{ST3}. Because this problem has an innermost MIP structure, it cannot be transformed into a single-level form straightly through dualization or the use of Karush-Kuhn-Tucker (KKT) condition. 
Alternatively, we construct an equivalent trilevel formulation by separating the binary and continuous recourse variables:
\begin{eqnarray}
	&&\mathbf{SP_1}: \Psi(\bm{x},\bm{z})=\min_{\bm{u},\bm{H}_0 \in \mathcal{U}(\bm{x},\bm{z})} \max_{\bm{\gamma},\bm{\omega},\bm{y}\in \mathcal{Q}({\bm{z}},\bm{u},\bm{H}_0) } \bm{q}^T\bm{y} \label{IV-A-OBJ1} \\
	&&\quad \quad \quad \ =\min_{\bm{u},\bm{H}_0 \in \mathcal{U}(\bm{x},\bm{z})} \max_{\bm{\gamma},\bm{\omega}\in \Theta} \max_{\bm{y}\in\mathcal{Q}_r(\bm{u},\bm{H}_0,\bm{\gamma},\bm{\omega})} \bm{q}^T\bm{y} \label{IV-A-OBJ2}\\
	&&\Theta=\{\bm{\gamma},\bm{\omega}\in \{0,1\}^{r_1\times r_2}:\ \bm{J}_1\bm{\gamma}+\bm{J}_2\bm{\omega}\geq \bm{l}-\bm{T}{\bm{z}} \}  \label{IV-A-CONS1}\\
	&&\mathcal{Q}_r(\bm{u},\bm{H}_0,\bm{\gamma},\bm{\omega})=\{\bm{y}\in\mathbb{R}^p:  \label{IV-A-CONS2}\\
	&&\quad \quad \quad \quad \quad \quad  \quad \ \bm{W}\bm{\gamma}+\bm{F}\bm{y}\geq \bm{f}-\bm{G}{\bm{H}_0}\label{IV-A-CONS3}\\
	&&\quad \quad \quad \quad \quad \quad \quad \  \bm{H}\bm{y}+\bm{L}\bm{\omega}\bm{u}\geq \bm{h} \} \label{IV-A-CONS4}
\end{eqnarray}
where $\Theta$ defines an independent constraints set for the discrete variables $\bm{\gamma}$ and $\bm{\omega}$. By excluding $\Theta$, the original recourse constraints set $\mathcal{Q}$ can be modified as $\mathcal{Q}_r$ in \eqref{IV-A-CONS2}-\eqref{IV-A-CONS4}. So the innermost-level problem of $\mathbf{SP_1}$ becomes a linear program (LP). By leveraging its strong duality, $\mathbf{SP_1}$ can be equivalently transformed into the following  ``min-max-min'' problem:
\setlength{\arraycolsep}{-0.0em}
\begin{eqnarray}
	&&\min_{\bm{u},\bm{H}_0 \in \mathcal{U}(\bm{x},\bm{z})} \max_{\bm{\gamma},\bm{\omega}\in \Theta} \min_{\bm{\lambda},\bm{\pi}\in\Omega_D} \bm{\lambda}^T(\bm{f}-\bm{G}{\bm{H}_0}-\bm{W}\bm{\gamma}) \nonumber \\
	&&\quad\quad\quad\quad\quad\quad\quad\quad \quad\quad   +\bm{\pi}^T (\bm{h}-\bm{L}\bm{\omega}\bm{u}) \label{ETRI:OBJ}\\ 
	&&\quad \Omega_D=\{ \bm{\lambda},\bm{\pi}\in\mathbb{R}^{s_1\times s_2}_-: \label{ETRI:CONS1}\\
	&&\quad \quad \quad \quad \bm{F}^T \bm{\lambda}+\bm{H}^T \bm{\pi}=\bm{q}  \} \label{ETRI:CONS3}
\end{eqnarray}
where $\bm{\lambda}$, $\bm{\pi}$ are dual variables corresponding to \eqref{IV-A-CONS3} and \eqref{IV-A-CONS4}. Note that $\Omega_D$ is a fixed polyhedron independent of any decision variable. 

Due to the slackness of load serving variables ($\bm{pl}$, $\bm{ql}$) in energy balance constraints, the relatively complete recourse property holds for problem \eqref{ETRI:OBJ}-\eqref{ETRI:CONS3}. Hence, for any given $\bm{x}$ and $\bm{z}$, there exists an optimal solution ($\bm{\gamma}^*,\bm{\omega}^*$) with ($\bm{\lambda}^*,\bm{\pi}^*$) being an extreme point of $\Omega_D$. Let $\Pi$ be the full set of extreme points of $\Omega_D$ which is fixed and finite. Considering the feasible solution set of decision variables $(\bm{\gamma},\bm{\omega})$ are finite, we define $\mathcal{Y}=\{(\bm{\gamma}_1,\bm{\omega}_1),...,(\bm{\gamma}_{|\mathcal{Y}|},\bm{\omega}_{|\mathcal{Y}|})\}$ which contains all feasible decisions of $\bm{\gamma}$ and $\bm{\omega}$.  Let $\mathcal{D}=\{(\bm{\lambda}_1,\bm{\pi}_1),...,(\bm{\lambda}_{|\mathcal{Y}|},\bm{\pi}_{|\mathcal{Y}|}) \in \Pi^{|\mathcal{Y}|}\}$. Then for a given $\mathcal{D}$, the worst-case scenario out of $\mathcal{U}(\bm{x},\bm{z})$ can be characterized as the solution of the following MIP \cite{wang2023computing}:
\begin{eqnarray}
	&&\Psi(\bm{x},\bm{z},\mathcal{D})=\min \ \eta  \label{MMR2:OBJ}\\
	&&s.t. \ \eta \geq \bm{\lambda}^T_i(\bm{f}-\bm{G}\bm{H}_0-\bm{W}\bm{\gamma}_i)+\bm{\pi}^T_i (\bm{h}-\bm{L}\bm{\omega}_i\bm{u}), \nonumber\\  
	&&\quad\quad\quad\quad\quad\quad\quad\quad\quad\quad\quad\quad\quad\quad\quad\quad i=1,\ldots,|\mathcal{Y}|\label{MMR2:CONS1}\\
	&&\quad \ \bm{u},\bm{H}_0 \in \mathcal{U}(\bm{x},\bm{z}) \label{MMR2:CONS2}
\end{eqnarray}

In general, we cannot have any closed-form to characterize this MIP. Nevertheless, it has some nice properties that reduce it to its LP relaxation. 

\begin{lem}\label{lem_integral_U}
For any fixed $(\bm x, \bm z)$, the $\bf u$-portion of any extreme point of the continuous relaxation of  $\mathcal{U}(\bm{x},\bm{z})$, denoted by  $\mathcal{U}^r(\bm{x},\bm{z})$, is binary. 
\end{lem} 
 Lemma \ref{lem_integral_U} can be easily proven by showing that, for any LP defined  over $\mathcal{U}^r(\bm{x},\bm{z})$, there exists one optimal solution with this property due to the decoupling characteristic of $\bm{u}$ and $\bm{H}_0$, and the \emph{totally unimodular} property of \eqref{DDU1}-\eqref{DDU2}. Additionally, let $\mathcal{F}(\bm{x},\bm{z}, \bm{u},\bm{H}_0)$ denote the recourse value for given $(\bm{x},\bm{z}, \bm{u},\bm{H}_0)$. It can be seen easily that $\mathcal{F}(\bm{x},\bm{z}, \bm{u},\bm{H}_0)$ monotonically increases  with respect to $\bm u$ and $\bm H_0$. According to \cite{majthay1974quasi}, together with Lemma \ref{lem_integral_U}, a strong result is presented next for the min-max optimization problem in (51). 

\begin{prop}\label{prop_integral_U}
 Function $\mathcal{F}(\bm{x},\bm{z},\bm{u},\bm{H}_0)$ is quasi-concave with respect to  $(\bm{u},\bm{H}_0)$. Also, an optimal solution to the min-max optimization problem in \eqref{IV-A-OBJ1} defined on $\mathcal{U}^r(\bm{x},\bm{z})$ is also optimal to this problem defined on $\mathcal{U}(\bm{x},\bm{z})$.
\end{prop} 

According to \textbf{Proposition \ref{prop_integral_U}}, an  optimal solution of problem \eqref{IV-A-OBJ1} can be derived by solving optimization problem \eqref{MMR2:OBJ}-\eqref{MMR2:CONS1} defined on  $\mathcal{U}^r(\bm{x},\bm{z})$, which can be characterized by its KKT conditions (represented by $\mathcal{OU}(\bm{x},\bm{z},\mathcal{D})$), providing a projection onto the space of $\mathcal{U}(\bm{x},\bm{z})$ \cite{wang2023computing}. However, due to the massive cardinality of $\mathcal{Y}$, the scale of such linear program is so enormous that its computation is practically infeasible. Actually, most of the constraints in \eqref{MMR2:CONS1} constructed by the $(\bm{\gamma},\bm{\omega})$ pairs in $\mathcal{Y}$ are ineffective. 
Hence, we can further simplify it as below.

Let $\bm{\upsilon}$ and $\bm{\mu}$ represent the $(\bm{\gamma},\bm{\omega})$ pair and $(\bm{\lambda},\bm{\pi})$ pair,  respectively. Rather than enumerating all $\bm{\upsilon}$'s in $\mathcal{Y}$ to formulate $\Psi(\bm{x},\bm{z},\mathcal{D})$, we could only consider a subset of $(\bm{\upsilon},\bm{\mu})$ pairs to construct a relaxed one. Let $\widehat{(\bm{\upsilon},\bm{\mu})}=\left\lbrace (\bm{\upsilon}_1,\bm{\mu}_1),...,(\bm{\upsilon}_{|\hat{\mathcal{Y}}|},\bm{\mu}_{|\hat{\mathcal{Y}}|})\right\rbrace $ denote such a subset with $\hat{\mathcal{Y}}=\left\lbrace \bm{\upsilon}_1,...,\bm{\upsilon}_{|\hat{\mathcal{Y}}|}\right\rbrace $ and associated set $\hat{\mathcal{D}}=\left\lbrace \bm{\mu}_1,...,\bm{\mu}_{|\hat{\mathcal{Y}}|}\right\rbrace$. Note that $\bm{\upsilon}$'s in $\hat{\mathcal{Y}}$ are different while some $\hat{\bm{\mu}}$'s in $\hat{\mathcal{D}}$ can be the same. 
The scale of problem \eqref{MMR2:OBJ} defined on $\widehat{(\bm{\upsilon},\bm{\mu})}$ could be much smaller than that of  \eqref{MMR2:OBJ}-\eqref{MMR2:CONS2} and more practical to formulate the optimal uncertainty set, denoted by $\mathcal{OU}(\bm{x},\bm{z},\widehat{(\bm{\upsilon},\bm{\mu})})$. 

Although $\mathcal{U}(\bm{x},\bm{z})$ is not fixed, $\mathcal{OU}$ actually provides a parametric approach to characterize the nontrivial scenarios in $\mathcal{U}(\bm{x},\bm{z})$. Furthermore, these scenarios, which are optimal to some $\widehat{(\bm{\upsilon},\bm{\mu})}$ sets, can change with $\bm{x}$ and $\bm{z}$.

\subsection{Equivalent Single-Level Reformulation}
As mentioned above, $\cal Y$ and $\Pi$ are fixed and finite, as well as independent of first-stage decisions $\bm{x}$ and $\bm{z}$. Therefore, through the complete enumeration of $\widehat{(\bm{\upsilon},\bm{\mu})}$, $\mathbf{DDU-RRO}$ can be transformed into an equivalent single-level formulation, which is the foundation of our solution method. 

\begin{prop}
	\label{prop:2}
	Let $2^\mathcal{Y}$ denote the power set of $\mathcal{Y}$. For fixed $\bm{x}$ and $\bm{z}$, we have
	\begin{eqnarray}
		&&\quad \min_{\bm{u},\bm{H}_0\in \mathcal{U}(\bm{x},\bm{z})} \max_{\bm{\gamma},\bm{\omega},\bm{y}\in \mathcal{Q}(\bm{z},\bm{u},\bm{H}_0)} \bm{q}^T\bm{y} \nonumber \\ 
		&&=\min_{\bm{u},\bm{H}_0\in \mathcal{U}^*(\bm{x},\bm{z})} \max_{\bm{\gamma},\bm{\omega},\bm{y}\in \mathcal{Q}(\bm{z},\bm{u},\bm{H}_0) } \bm{q}^T\bm{y} \label{lem1}
	\end{eqnarray}
	where $\mathcal{U}^*(\bm{x},\bm{z})=$
	
	$\left\lbrace \bm{u},\bm{H}_0:\bm{u},\bm{H}_0\in \mathcal{OU}(\bm{x},\bm{z},\widehat{(\bm{\upsilon},\bm{\mu})}), \forall \widehat{(\bm{\upsilon},\bm{\mu})} \in 2^\mathcal{Y} \times \Pi^{\hat{|\mathcal{Y}|}} \right\rbrace $.
\end{prop}
The detailed proof can be found in \cite{wang2023computing}.

We note that $\mathcal{U}^*(\bm{x},\bm{z})$ is the union of a finite number of projections of $\mathcal{OU}$ sets. As a consequence, we can obtain the following single-level formulation.

\begin{thm}
	$\mathbf{DDU-RRO}$ is equivalent to the following single-level problem:
	\begin{eqnarray}
		&&{\bf DDU-Single}: \ \Gamma=\min  \ \bm{c}^T\bm{x}+ \bm{d}^T\bm{z} \label{SROHM:OBJ}\\
		&&s.t. \ \bm{x},\bm{z} \in \mathcal{X}   \label{SROHM:CONS1}\\
		&&\bm{q}^T\bm{y}_l \geq \hat{\Upsilon},\quad l=1,...,N \label{SROHM:RC}\\
		&&\bm{W}\bm{\gamma}_l+\bm{F}\bm{y}_l\geq \bm{f}-\bm{G}\bm{H}_0^l,\quad l=1,...,N \label{SROHM:CONS2}\\
		&&\bm{H}\bm{y}_l+\bm{L}\bm{\omega}_l\bm{u}_l\geq \bm{h},\quad l=1,...,N \label{SROHM:CONS3}\\
		&&\bm{J}_1\bm{\gamma}_l+\bm{J}_2\bm{\omega}_l\geq \bm{l}-\bm{T}\bm{z} ,\quad l=1,...,N \label{SROHM:CONS4} \\
		&&\bm{u}_l,\bm{H}_0^l\in \mathcal{OU}(\bm{x},\bm{z},\widehat{(\bm{\upsilon},\bm{\mu})}_l),\quad l=1,...,N \label{SROHM:OU}\\
		&&\bm{\gamma}_l,\bm{\omega}_l\in \{0,1\}^{r_1\times r_2}, \bm{y}_l\in  \mathbb{R}^p, \quad l=1,...,N \label{SROHM:var}
	\end{eqnarray}
	where $N$ is the cardinality of $2^\mathcal{Y} \times \Pi^{\hat{|\mathcal{Y}|}} $.
\end{thm}

Through the incomplete enumeration of $\widehat{(\bm{\upsilon},\bm{\mu})}$ whose set is denoted by $\overline{(\hat{\mathcal{Y}},\hat{\cal D})}$, we can formulate a relaxed problem of ${\bf DDU-RRO}$, and the $\overline{(\hat{\mathcal{Y}},\hat{\cal D})}$ set can be expanded through a well-designed computation oracle. This idea motivates us to develop the N-PC\&CG that includes outer- and inner- loop algorithms. 

\subsection{Outer-Loop Parametric C\&CG Algorithm}
\label{sec:4B}
Considering that $\mathbf{SP_1}$ can be exactly computed for any given $\left(\bm{x},\bm{z}\right)$ (as demonstrated in Section \ref{sec:4A}), its dual solution helps characterize the nontrivial scenarios in a parametric way. Then, the master problem ($\mathbf{MP_1}$), would be strengthened by adding the corresponding $\mathcal{OU}$. 
Accordingly, the customized computation procedures of our outer-loop algorithm are presented as in the following.

\noindent $\blacksquare$ \textbf{Algorithm 1:} Outer-Loop Parametric C\&CG
\begin{description}
	\item [$\rm{Step \ 1.1}$] $\quad $ Set a flag variable $\Lambda=1$; Initialize $n=1$ and set $\overline{(\hat{\mathcal{Y}},\hat{\cal D})}=\emptyset$.
	\item [$\rm{Step \ 1.2}$] $\quad$ Compute the outer-loop master problem $\mathbf{MP}_1^n$.
	\begin{eqnarray}
		&&\mathbf{MP}_1^n: \ \Gamma=\min  \ \bm{c}^T\bm{x}+ \bm{d}^T\bm{z} \label{MP1:OBJ}\\
		&&s.t. \ \bm{x},\bm{z} \in \mathcal{X}   \label{MP1:CONS1}\\			
		&&\bm{q}^T\bm{y}_l \geq \hat{\Upsilon},\quad l=1,...,n-1 \label{MP1:RC}\\	
		&&\bm{W}\bm{\gamma}_l+\bm{F}\bm{y}_l\geq \bm{f}-\bm{G}\bm{H}_0^l,\quad l=1,...,n-1 \label{MP1:CONS2}\\
		&&\bm{H}\bm{y}_l+\bm{L}\bm{\omega}_l\bm{u}_l\geq \bm{h},\quad l=1,...,n-1 \label{MP1:CONS3}\\
		&&\bm{J}_1\bm{\gamma_l}+\bm{J}_2\bm{\omega}_l\geq \bm{l}-\bm{T}\bm{z} ,\quad l=1,...,n-1 \label{MP1:CONS4} \\
		&&\bm{u}_l,\bm{H}_0^l\in \mathcal{OU}(\bm{x},\bm{z},\widehat{(\bm{\upsilon},\bm{\mu})}_l),\quad l=1,...,n-1 \label{MP1:OU}\\
		&&\bm{\gamma}_l,\bm{\omega}_l\in \{0,1\}^{r_1\times r_2}, \bm{y}_l\in  \mathbb{R}^p, \quad l=1,...,n-1 \label{MP1:var}
	\end{eqnarray}
	where $n$ represents the index of iteration. If $\mathbf{MP}_1^n$ is infeasible, then terminate and report the infeasibility of $\mathbf{DDU-RRO}$. Otherwise, derive solution ($\hat{\bm{x}}_n$, $\hat{\bm{z}}_n$)  and the current optimal value $\hat{\Gamma}$.
	\item [$\rm{Step \ 1.3}$] $\quad$ Compute the subproblem $\mathbf{SP}_1^n$ for given ($\hat{\bm{x}}_n$, $\hat{\bm{z}}_n$) to attain critical discrete recourse decisions and corresponding extreme points which form the set $\widehat{(\bm{\upsilon},\bm{\mu})}_n$, and optimal value $\Psi(\hat{\bm{x}}_n, \hat{\bm{z}}_n)$. Update $\overline{(\hat{\mathcal{Y}},\hat{\cal D})}=\overline{(\hat{\mathcal{Y}},\hat{\cal D})}\bigcup \{ \widehat{(\bm{\upsilon},{\bm{\mu}})}_n \}$.
	\item [$\rm{Step \ 1.4}$] $\quad$ If $\Psi(\hat{\bm{x}}_{n},\hat{\bm{z}}_n) \geq \hat{\Upsilon}$, set $\Lambda=0$ to indicate the feasibility of $\min-\max$ resilience constraint. Then, terminate the outer-loop algorithm and report the optimal solution $(\bm{x}^*,\bm{z}^*)=(\hat{\bm{x}}_n,\hat{\bm{z}}_n)$. Otherwise, create new variables $(\bm{\gamma}_n, \bm{\omega}_n, \bm{y}_n, \bm{u}_n,\bm{H}_0^n)$ and add the following cutting set to $\mathbf{MP}_1^{n+1}$.
	\setlength{\arraycolsep}{-0.4em}
	\begin{eqnarray}
		&&\bm{q}^T\bm{y_{n}} \geq \hat{\Upsilon}\label{Cut1}\\
		&&\bm{W}\bm{\gamma}_n+\bm{F}\bm{y}_n\geq \bm{f}-\bm{G}\bm{H}_0^n \label{Cut2}\\
		&&\bm{H}\bm{y}_n+\bm{L}\bm{\omega}_n\bm{u}_n\geq \bm{h} \label{Cut3}\\
		&&\bm{J}_1\bm{\gamma}_n+\bm{J}_2\bm{\omega}_n\geq \bm{l}-\bm{T}\bm{z}  \label{Cut4} \\
		&&\bm{u}_n,\bm{H}_0^n\in \mathcal{OU}(\bm{x},\bm{z},\widehat{(\bm{\upsilon},\bm{\mu})}_n) \label{Cut5} \\
		&&\bm{\gamma}_n,\bm{\omega}_n\in \{0,1\}^{r_1\times r_2}, \bm{y}_n\in  \mathbb{R}^p \label{Cut6}
	\end{eqnarray}
	Then set $n\gets {n+1}$ and go ot $\rm{Step \ 2}$.
\end{description}

\begin{remark}
\label{rmk:4}
Notice again that $2^\mathcal{Y}$ and $\Pi$ are fixed and finite sets, so the number of $\widehat{(\bm{\upsilon},{\bm{\mu}})}$ is finite. Hence, the N-PC\&CG will terminate by either reporting the infeasibility of $\mathbf{DDU-RRO}$ or converging to the global optimal value with an exact solution after a finite number of iterations.
\end{remark} \vspace{-10pt}

\subsection{Inner-Loop C\&CG Algorithm}
\label{sec:4C}
According to Section \ref{sec:4A}, the dual solution of $\mathbf{SP}_1$ (that returns $\widehat{(\bm{\upsilon},{\bm{\mu}})}$) can be exactly attained by computing its trilevel reformulation as in \eqref{ETRI:OBJ}-\eqref{ETRI:CONS3}. Then, the C\&CG procedures for solving bi-level MIP  can be applied as follows. 

\noindent $\blacksquare$ \textbf{Algorithm 2:} Inner-Loop C\&CG
\begin{description}
	\item [$\rm{Step \ 2.1}$] $\quad$ Initialize $LB=-\infty$, $UB=+\infty$, $j=1$. Set $\hat{\mathcal{Y}}=\emptyset$.
	\item [$\rm{Step \ 2.2}$] $\quad$ Solve the inner-loop master problem $\mathbf{MP}_2^j$.
	\setlength{\arraycolsep}{-0.45em}
	\begin{eqnarray}
		&&{\bf MP}_2^j: \ \Psi(\hat{\bm{x}},\hat{\bm{z}})=\min \ \eta  \label{MP2:OBJ}\\
		&&s.t. \ \eta \geq \bm{\lambda}^T_i(\bm{f}-\bm{G}{\bm{H}_0}-\bm{W}\hat{\bm{\gamma}}_i)+\bm{\pi}^T_i (\bm{h}-\bm{L}\hat{\bm{\omega}}_i\bm{u}), \nonumber\\  
		&&\quad\quad\quad\quad\quad\quad\quad\quad\quad\quad\quad\quad\quad\quad i=1,\ldots,j-1\label{MP2:CONS1}\\
		&&\bm{F}^T \bm{\lambda}_i+\bm{H}^T \bm{\pi}_i=\bm{q}, \quad i=1,\ldots,j-1  \label{MP2:CONS3}\\
		&&\bm{u},\bm{H}_{0}\in \mathcal{U}(\hat{\bm{x}},\hat{\bm{z}}), \ \bm{\lambda}_i,\bm{\pi}_i\in\mathbb{R}^{s_1\times s_2}_-,  i=1,\ldots,j-1 \label{MP2:CONS4}
	\end{eqnarray}
	By computing ${\bf MP}_2^j$, we can derive optimal solution $\bm{\hat{u}}_j$, $\hat{\bm{H}}_0^j$ and $\{(\hat{\bm{\lambda}}_1,\hat{\bm{\pi}}_1),...,(\hat{\bm{\lambda}}_{j-1},\hat{\bm{\pi}}_{j-1})\}$. Update the lower bound of ${\bf SP}_1$, i.e., $LB=\hat{\eta}_j$;
	\item [$\rm{Step \ 2.3}$] $\quad$ Solve the subproblem $\mathbf{SP}_2^j$ for given $\hat{\bm{u}}_{j}$ and $\hat{\bm{H}}_0^j$.
	\begin{eqnarray}
		&&{\bf SP}_2^j: \ \max \ \bm{q}^T\bm{y} \label{SP2:OBJ}\\
		&&s.t. \ \bm{W}\bm{\gamma}+\bm{F}\bm{y}\geq \bm{f}-\bm{G}\hat{\bm{H}}_0^j \label{SP2:CONS1}\\
		&&\bm{H}\bm{y}+\bm{L}\bm{\omega}\hat{\bm{u}}_j\geq \bm{h} \label{SP2:CONS2}\\
		&&\bm{J}_1\bm{\gamma}+\bm{J}_2\bm{\omega}\geq \bm{l}-\bm{T}\hat{\bm{z}} \label{SP2:CONS3}\\
		&&\bm{\gamma},\bm{\omega}\in \{0,1\}^{r_1\times r_2},\bm{y}\in \mathbb{R}^p \label{SP2:CONS4}
	\end{eqnarray}
	Get solution $(\hat{\bm{\gamma}}_j,\hat{\bm{\omega}}_j,\hat{\bm{y}}_j)$  and update the upper bound of ${\bf SP}_1$,  i.e., $UB=\min\left\lbrace UB, \bm{q}^T\hat{\bm{y}}_j \right\rbrace$;
	\item [$\rm{Step \ 2.4}$] $\quad$ If $\frac{UB-LB}{LB}\leq \varepsilon$, terminate the inner-loop algorithm, return the optimal value $\Psi(\hat{\bm{x}},\hat{\bm{z}})$ and $\widehat{(\bm{\upsilon},\bm{\mu})}=\{ (\hat{\bm{\upsilon}}_1,\hat{\bm{\mu}}_1),..., 
	(\hat{\bm{\upsilon}}_{j-1},\hat{\bm{\bm{\mu}}}_{j-1}) \}$.
	Otherwise, update $\hat{\mathcal{Y}}=\hat{\mathcal{Y}} \bigcup \{  \hat{\bm{\upsilon}}_j \}$, create new variables $(\bm{\lambda}_{j}, {\bm{\pi}}_{j})$ and add the following constraints to $\mathbf{MP}_2^{j+1}$.
	\setlength{\arraycolsep}{-0.8em}
	\begin{eqnarray}
		&&\eta \geq \bm{\lambda}^T_j(\bm{f}-\bm{G}\bm{H}_0-\bm{W}\hat{\bm{\gamma}}_j)+\bm{\pi}^T_j (\bm{h}-\bm{L}\hat{\bm{\omega}}_j\bm{u}) \label{SCUT1}\\
		&&\bm{F}^T \bm{\lambda}_j+\bm{H}^T \bm{\pi}_j=\bm{q}  \label{SCUT2}\\	
		&&\bm{\lambda}_j,\bm{\mu}_j\in\mathbb{R}^{s_1\times s_2}_+ \label{SCUT3}
	\end{eqnarray}
	Set $j\gets {j+1}$ and go to $\rm{Step \ 2.2}$.
	\item [$\rm{Step \ 2.5}$]  $\quad$ Solve the correction problem ($\mathbf{CP}$) as below.
	\begin{eqnarray}
		&&{\bf CP}: \ \mathcal{C}=\max \ \bm{q}^T\bm{y} \label{CP:OBJ}\\
		&&s.t. \ \bm{W}\bm{\gamma}+\bm{F}\bm{y}\geq \bm{f}-\bm{G}\bm{H}_0 \label{CP:CONS1}\\
		&&\bm{H}\bm{y}+\bm{L}\bm{\omega}{\bm{u}}\geq \bm{h} \label{CP:CONS2}\\
		&&\bm{J}_1\bm{\gamma}+\bm{J}_2\bm{\omega}\geq \bm{l}-\bm{T}\hat{\bm{z}} \label{CP:CONS3}\\
		&&\bm{u},\bm{H}_0\in \mathcal{OU}(\hat{\bm{x}},\hat{\bm{z}},\widehat{(\bm{\upsilon},\bm{\mu})})  \label{CP:CONS4} \\
		&&\bm{\gamma},\bm{\omega}\in \{0,1\}^{r_1\times r_2},\bm{y}\in \mathbb{R}^p \label{CP:CONS5}
	\end{eqnarray}
	Derive the optimal value $\mathcal{C}$, optimal solution $(\hat{\bm{u}},\hat{\bm{H}}_0,\hat{\bm{\gamma}},\hat{\bm{\omega}})$. The corresponding extreme points $(\hat{\bm{\lambda}},\hat{\bm{\pi}})$ can be obtained from the inner-most problem of \eqref{ETRI:OBJ}-\eqref{ETRI:CONS3} for given $(\hat{\bm{x}},\hat{\bm{z}},\hat{\bm{u}},\hat{\bm{H}}_0,\hat{\bm{\gamma}},\hat{\bm{\omega}})$.
	\item[$\rm{Step \ 2.6}$] $\quad$ if $\mathcal{C}>UB$, update $\widehat{(\bm{\upsilon},{\bm{\mu}})}=\widehat{(\bm{\upsilon},{\bm{\mu}})} \bigcup \{(\hat{\bm{\upsilon}},\hat{\bm{\mu}})\}$ and go to $\rm{Step \ 2.5}$. Otherwise, terminate the inner-loop algorithm, return the optimal value $\Psi(\hat{\bm{x}},\hat{\bm{z}})$ and $\widehat{(\bm{\upsilon},{\bm{\mu}})}$.

\end{description}

\begin{remark}
\label{rmk:5}
Based on \eqref{IV-A-CONS1}, for given $\bm{x}$ and $\bm{z}$, $\Theta$ is a fixed and finite set with respect to $(\bm{\gamma},\bm{\omega})$. It follows that the inner-loop C\&CG can converge and report $\widehat{(\bm{\upsilon},\bm{\mu})}$ in finite steps. 	
\end{remark}

\subsection{Enhancement strategies}
Note that the core step of the N-PC\&CG procedure is to formulate critical $\mathcal{OU}$ sets and generate primal cuts to strengthen the outer master problem. However, due to the complexity of power networks, \eqref{DDU1}-\eqref{DDU2} often has multiple optimal solutions, rendering $\mathcal{OU}$ a non-singleton set \cite{ma2025proactive}. With this issue, extra iterations may be involved, leading to a longer computational time. Alleviating the issue of multiple solutions of $\mathcal{OU}$ could help improve the solution efficiency. Additionally, we believe that the recourse decisions in the $\mathbf{MP_1}$ contain valuable information, which can be utilized for the inner-loop C\&CG. These two ideas motivates us to develop the following enhancement strategies. 
	
\emph{1) Inclusion of $\mathcal{OU}'(\bm{x})$:} Our first enhancement is to construct an additional optimal solution set, i.e., $\mathcal{OU}'(\bm{x})$, for uncertain variables $\bm{u}$. Assuming that for given $\bm{x}$ and $\bm{z}$, $\mathbf{SP_1}$ derives the worst-case scenario $\bm{u}^*$ using $r$ iterations. Then, we define the following auxiliary problem corresponding to $\mathcal{OU}'(\bm{x})$. 
\begin{equation}
	\min \{ \bm{a}^T \bm{u}+\bm{\vartheta}^T \bm{u}: \ s.t. \ Eqs. (4)-(5) \} \label{Enhancement1}
\end{equation}
where $\bm{a}$ and $\bm{\vartheta}$ are the primary coefficient vector and the modified term, respectively. Specifically, let $\bm{a}$ equal $M_2$, with $M_2$ being a very large constant whenever $\bm{u}^*_{ij}=0$. As mentioned in \cite{ma2025proactive}, such setting ensures that $\mathcal{OU}'(\bm{x})$ is a singleton that only contains $\bm{u}^*$, as long as the set of power lines being hardened does not overlap the lines corresponding to zero entries of $\bm{u}^*$. Thus, $\mathcal{OU}'(\bm{x})$ can precisely capture the faulting information of $\bm{u}^*$, which is equivalent to the $\bm{u}$-portion of corresponding $\mathcal{OU}(\bm{x},\bm{z})$. Additionally, $\bm{\vartheta}$ contributes to reducing the optimal solution of $\mathcal{OU}'(\bm{x})$ to a singleton or a much small subset of the original $\mathcal{OU}$ set. For the sake of illustration, it could be defined as $\sum_{i=1}^{r} (1-\bm{u}_i)$ where $\bm{u}_i$ is the optimal solution of $\mathbf{MP_2}$ in the inner-loop C\&CG. To a certain extent, it can reflect the importance of other lines except the damaged lines in $\bm{u}^*$. Consequently, the application of $\mathcal{OU}'(\bm{x})$ provides an effective way to identify the lines that are more adversarial to network hardening, and thus improving the convergence performance of N-PC\&CG (see our previous studies in \cite{ma2025proactive} for more details).

\emph{2) Warm start for inner-loop C\&CG:} Another important enhancement is to strengthen the inner-loop C\&CG by including prior scenarios and their corresponding constraints in $\mathbf{MP_2}$. In particular, for the $n$-th outer iteration, we can initialize the inner-loop algorithm using a set $\hat{\cal Y}=\{\bm{\upsilon}_1,...,\bm{\upsilon}_{n-1}\}$, where each $\bm{\upsilon}_i$ is derived from the binary recourse decisions in $\mathbf{MP}_1^n$. The incorporation of these cutting planes enables the inner-loop C\&CG to better converge to the worst-case contingency scenario $\bm{\upsilon}^*$, thereby reducing the computational time.

\section{Numerical Results}
\label{Sec:Case}

The effectiveness and scalability of proposed robust co-planning model and N-PC\&CG algorithm are validated on 14-bus test system and 56-bus real-world distribution network. The renting cost of MHERs is considered to be 2\% of their capital expenditures (CAPEXs) \cite{wang2023resilience}. The initial hydrogen storage level of each MHER is no less than 70\% ($\sigma_1$) of its maximum capacity. Particularly, the total storage level of MHERs that are charged more with higher rental costs is required to exceed 90\% ($\sigma_2$). We assume that the structural hardening is conducted on those normally connected lines, with an unit hardening cost as \$120,000/km \cite{Zhang2021TSG}. Note also that the network hardening is essentially a long-term investment action. To be consistent with the short-term renting cost of MHERs (incurred by one-time emergency response), we use the average hardening cost per event according to the statistical occurrence frequency of disastrous instances.  

Our tests are performed in MATLAB R2021a on a desktop computer with Intel Core i9-10900K 3.70GHz processor (10 cores) and 128GB memory. All the MPs and SPs in N-PC\&CG algorithm are computed using Gurobi 11.0.

\subsection{14-Bus Test Distribution Network}

\begin{figure}[h]
	\centering
	\includegraphics[width=7.5cm,height=4.2cm]{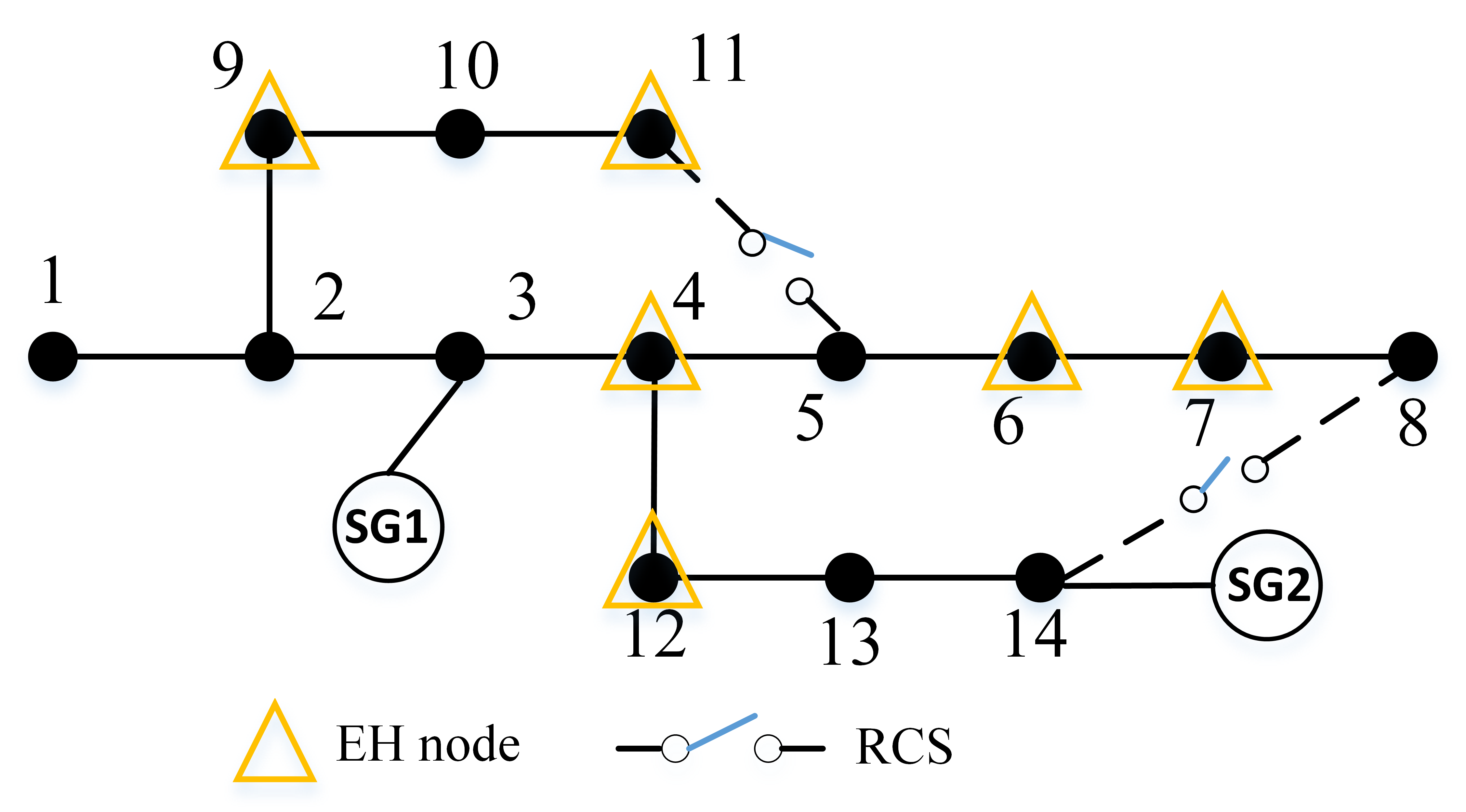}
	\caption{14-Bus Test Power Distribution Network} \vspace{-10pt}
	\label{fig:PND} 
\end{figure}

As exhibited in Fig. \ref{fig:PND}, the 14-bus test network includes 6 EH nodes, 6 MHERs and 2 SGs. The installed capacity of SG1 and SG2 are 200kW and 250kW, respectively. All these MHERs are categorized into three types, whose techno-economical parameters are presented in Table \ref{tab:mher}. In our test network, two distribution lines (L5-11, L8-14) are equipped with tie switches (normally opened and assumed to be invulnerable), while the rest lines with sectionalizing switches (normally closed). 

\renewcommand\arraystretch{1.00}
\begin{table}
	\centering
	\caption{Techno-Economical Parameters of MHERs} \vspace{-5pt}
	\setlength{\tabcolsep}{1.00mm}
	\scalebox{0.85}{
		\begin{tabular}{ccccccccc}
			\toprule
			Type &   Number    &   $\overline{H}$  &   $\overline{P}$  & $\overline{Q}$    &  $\xi$    &  $\eta$    &  $c^H$    & $m^{tp}$   \\
			\midrule
			1     & 1/4  & 30kg  & 200kW & 250kVar      & \multirow{3}{*}{38.9kWh/kg} & \multirow{3}{*}{52\%} & 8.3/9.96k\$  & \multirow{3}{*}{1kg/h} \\
			2     & 2/5     & 45kg & 200kW & 300kVar       &       &   & 8.45/10.14k\$  &  \\
			3     & 3/6   & 60kg & 300kW & 400kVar       &       &   & 12.6/15.12k\$  &  \\
			\bottomrule
	\end{tabular}}
	\label{tab:mher}%
\end{table}

\subsubsection{Effectiveness of Resilient Co-Planning Strategy}
To verify the resilience enhancement effects of the proposed co-planning strategy, we perform comparative case studies as in the following:
\begin{itemize}
	\item[-] \textbf{Case 1:} Only with distribution network hardening.
	\item[-] \textbf{Case 2:} Only with MHERs' scheduling.
	\item[-] \textbf{Case 3:} Co-planning of network hardening and MHERs' scheduling.
\end{itemize}

We consider $N-5$ contingencies with three target levels of load survivability, i.e., $\hat{\Upsilon}=$ 80\%, 90\% and 95\%. In all the test cases, the topology reconfiguration can be implemented as an operational corrective measure. Also, for mobile resource scheduling, a maximum number of 2 MHERs are allowed to park in every EH node. By computing our $\mathbf{DDU-RRO}$ model, the test results of different cases are recorded in Table \ref{tab:test1}, including the optimal planning schemes, worst-case scenarios and system reinforcement costs (SRC). Note that the planning schemes and worst-case scenarios are expressed in a concise way. For example, L1-2 indicates the proactive hardening of line 1-2, M2N4 represents the deployment of MHER2 at EH node 4, and M3/0.7 means that the available storage level of MHER3 is 70\% of the nominal capacity.

\renewcommand\arraystretch{0.8}
\begin{table}
	\centering
	\caption{Results of Case 1-3 under Different Resilience Constraints} \vspace{-5pt}
	\setlength{\tabcolsep}{1.5mm}
	\scalebox{0.85}{
		\begin{tabular}{ccccc}
			\toprule
			Case & $\hat{\Upsilon}$ & Optimal Planning Scheme & Worst-Case Scenario & SRC/k\$  \\			
			\midrule
			\vspace{5pt}
			& 0.8  & \parbox{2.5cm}{L1-2, 2-3, 3-4, 4-5, \\5-6, 6-7} &  L2-9, 10-11, 4-12, 12-13, 13-14  & 91.4 \\
			\vspace{5pt}
			Case 1 & 0.90 & \parbox{2.5cm}{L1-2, 4-5, 5-6, 6-7, \\ 2-9, 9-10, 10-11} & L3-4, 4-12, 7-8, 12-13, 13-14 & 107.2 \\
			& 0.95 & \parbox{2.5cm}{L1-2, 4-5, 5-6, 6-7, 2-\\9, 9-10, 10-11, 13-14} & L2-3, 3-4, 4-12, 7-8, 12-13 & 125.6 \\
			\midrule 
			\vspace{5pt}
			& 0.80 & M3N7, M5N12 & \parbox{3cm}{ L2-9, 3-4, 4-5, 6-7, 12-13; \\ M3/0.7, M5/0.9} & 22.74 \\
			\vspace{5pt}
			Case 2 & 0.90 & / & / & / \\
			& 0.95 & / & / & /\\ 
			\midrule
			\vspace{5pt}
			& 0.80 & M3N7, M5N12 &\parbox{3.1cm}{\centering L2-9, 3-4, 4-5, 6-7, 12-13; \\ M3/0.7, M5/0.9}& 22.74 \\
			\vspace{5pt}
			Case 3 & 0.90 & \parbox{2.5cm}{\centering L2-9; M2N7,\\M4N12, M5N11} &\parbox{3.7cm}{\centering L1-2, 2-3, 10-11, 12-13, 13-14;\\ M2/0.7,M4/0.75,M5/1.0 } & 48.35 \\
			& 0.95 & \parbox{2.8cm}{\centering L1-2, 9-10, 12-13; \\M1N7,M2N4,M4N11,M5N12} &\parbox{3cm}{\centering L3-4, 4-5, 4-12, 5-6, 6-7;\\ M1/0.7,M2/0.7,M4/0.75,M5/1.0 } & 78.05 \\
			\bottomrule	  
	\end{tabular}}
	\label{tab:test1}
\end{table} 

\textit{$\bullet$ Case 1}:
As can be seen in Table \ref{tab:test1}, when $\hat{\Upsilon}$=80\%, 6 lines are hardened including the substation feeder. Affected by this hardening scheme, the worst-case scenario is identified as the outages of L2-9, L4-12, L10-11, L12-13, and L13-14. Then, at the post-event stage, the topology reconfiguration is performed to restore the power demands of N10 and N11 by closing the RCS on L5-11, while N14 is powered by the stationary generator SG2. But due to the damage of L2-9, L4-12, L10-11, and L13-14, the islanding area comprising of N9, N10, N12, and N13 are without any power sources. When $\hat{\Upsilon}$ increases from 80\% to 90\%, the protection of L2-3 and L3-4 are strategically abandoned while L2-9, L9-10, and L10-11 are reinforced. This adjustment raises the SRC from 91.4k\$ to 107.2k\$. Corresponding to the adjustment of hardening decisions, two critical damage spots changes (from L2-9, L10-11 to L3-4 and L7-8), while the other three remains. The protection of L2-9, L9-10, and L10-11 not only preserves the power supply of N9 and N10, but also provides power for vulnerable area N4-N7 by closing the RCS on L5-11. However, N12 and N13 are still isolated from all the power sources, which becomes the major cause of load curtailment. When $\hat{\Upsilon}$ further increases to 95\%, due to the immobility of SG1 and SG2, we may have to reinforce more distribution lines. The additional protection of L13-14 mitigates the loss of power risks of N13, leaving N12 as the only isolated site. As a result, the total SRC raises to 125.6k\$, which pays largely for the strategic hardening of 8 distribution lines. 

\textit{$\bullet$ Case 2}:
We only consider the proactive allocation and post-event scheduling of MHERs as the resilience enhancement measure. As shown in Table \ref{tab:test1}, such strategy could be infeasible under serious $N-k$ power contingencies. When $\hat{\Upsilon}=80\%$, MHER3 and MHER5 are proactively deployed at N7 and N12, which could be with 70\% and 90\% of the nominal storage capacity under the worst-case scenario. The worst contingency instance is identified as the damage of L2-9, L3-4, L4-5, L6-7, and L12-13, which partitions the PDN into 6 parts including four islands without any power sources. Then, at the post-event stage, the RCS at L8-14 are closed that connects N7, N8 to SG2. Besides, MHER3 is re-routed from N7 to N9, which coordinates with the connecting of L5-11 to energize the islanding region including N5, N6, and N9-11. Different from Case 1 that relies on stationary generators, the MHERs can be flexibly dispensed to any EH node with restorative needs, which remarkably increases the cost efficiency of deploying disaster-relief resources. Compared to Case 1, the SRC is reduced from 91.4k\$ to 22.74k\$ (decreasing by 75.1\%).  When $\hat{\Upsilon}\geq 90\%$, however, we cannot find any feasible robust solution by only using MHERs and without line hardening. A possible reason is that the MHERs is incapable to supply those non-EH nodes due to practical restrictions. These observations reveal the importance of structural reinforcement for addressing the high-standard resilience requirement.  

\textit{$\bullet$ Case 3}:
In this case, the network hardening and MHERs' scheduling are co-optimized. When $\hat{\Upsilon}=80\%$, the most economical scheme for resilience enhancement is only resorting to MHERs' scheduling. So we derive exactly the same planning scheme to Case 2. As $\hat{\Upsilon}$ increases to 90\%, L2-9 is hardened along with the deployment of MHER2, 4, 5 at N7, N12, and N11, respectively. Under the worst-case scenario, the residual storage level of MHER2 reaches its lower limit as 70\%. But due to the decision dependence of uncertain storage capacity, as in \eqref{DDU3}, the available storage levels of MHER4 and MHER5 are identified as 75\% and 100\% respectively. Note also that the substation feeder L1-2 as well as L2-3 and L10-11 could be destroyed, which creates an islanding region including N2 and N9-10 (with the pre-hardening of L2-9). To satisfy the load demands, MHER4 at N12 is scheduled to N9. On the other hand, by closing the RCSs at L5-11 and L8-14, the rest electric nodes excepting N13 are all connected, which can be powered by MHER 4-5 and SG 1-2. Compared to those in Case 1, our co-planning solution helps reduce the SRC by 47.0\%. With  $\hat{\Upsilon}=95\%$, L1-2, 9-10 and L12-13 are hardened, while MHER1 is additionally scheduled based on the planning scheme when $\hat{\Upsilon}$ = 90\%. Note that adequate MHERs can ensure that each island is connected to at least one MHER. To achieve the high-level resiliency target, it requires not only the protection of critical distribution lines, but also the reliable power supply for unintentional islands under an acceptable cost. 

\subsubsection{Influence of Damage Severity Levels}
The proposed resilience co-planning strategy is applied under different network damage levels (quantified by parameter $k$).  The target of load survivability is fixed as $\hat{\Upsilon}=90\%$, along with other conditions same as Case 3. The test results with $k=3, 4, 5, 6$ are presented in Table \ref{tab:test2}. Under $N-3$ instance, MHER3 and MHER5 are proactively allocated at N12 and N7, respectively. The distribution lines L2-9, L3-4, and L10-11 are damaged under the worst-case scenario, leading to the power outages of N9, 10, and 11. So, at the post-event stage, with the closing of RCS at L5-11, MHER5 is moved from N7 to N9 to support the power supply of this islanding region. By only dispatching two MHERs, it renders a rather low system cost as 18.59k\$. Under $N-4$ instance, it can still rely on mobile resources (i.e., MHER1, MHER2, and MHER5) to achieve the resiliency target, but the SRC is raised by 44.6\%. The corresponding worst-case scenario is to damage L9-10, L10-11, L12-13, and L13-14. By closing two RCSs, all the electric nodes excepting N10 and N13 are connected, and supplied by 2 stationary generators and 3 MHERs. Then, when the power contingencies become more severe (i.e., the $N-5$ and $N-6$ faults), we need both the network hardening and mobile resources scheduling to have sufficient self-restorative capability. Detailed analysis about the $N-5$ instance can be seen in Case 3. With $N-6$ faults, the substation feeder L1-2 is protected while 5 MHERs are utilized (MHER 1-5). Under the worst-case scenario, we lose L2-9, L4-12, L9-10, L10-11, L12-13, and 13-14, leading to four single-node islanding regions, i.e., N9, N10, N12, and N13. Specifically, N9 and N12 with V2G equipment can be powered by MHERs. The load curtailment occurs on N10 and N13. Compared to Case 3, the SRC increases from 48.35k\$ to 62.25k\$, which is practically acceptable considering such damage severity level. 

Moreover, we compare the optimization results of two test instances with $\hat{\Upsilon}=95\%$, $N-5$ and  $\hat{\Upsilon}=90\%$, $N-6$. An interesting observation is that the former one generates a higher hardening expense, which proves that our resilient planning strategy could be more sensitive to load survivability requirement than the damage severity. This result also confirms the scalability and good adaption of the proposed method to different contingency situations.

\renewcommand\arraystretch{1.2}
\begin{table}
	\centering
	\caption{Resilient Co-Planning Results under Different Damage Levels} \vspace{-5pt}
	\setlength{\tabcolsep}{1.5mm}
	\scalebox{0.85}{
		\begin{tabular}{cccc}
			\toprule
			$k$  & Resilient Co-Planning Scheme & Worst-Case Scenario & SRC/k\$  \\
			\midrule		
			3 & M3N12, M5N7  & L3-4, 2-9, 10-11 & 18.59  \\
			4 & M1N9, M2N11, M5N7  & L9-10, 10-11, 12-13, 13-14 & 26.89  \\	
			5 & L2-9; M2N7, M4N12, M5N11  & L1-2, 2-3, 10-11, 12-13, 13-14 & 48.35 \\
			6 & \parbox{2.8cm}{\centering L1-2; M1N11, M2N9,\\M3N12, M4N7, M5N6}  & L2-9, 4-12, 9-10, 10-11, 12-13, 13-14 & 62.05 \\	
			\bottomrule	  
	\end{tabular}}
	\label{tab:test2}
\end{table}  

\subsubsection{Results with Varying Renting Cost of MHERs}
We further investigate the impact of short-term renting cost of MHERs on the resilient co-planning results. Following the ideas in \cite{wang2023resilience}, we consider the ratio of MHERs' rental cost to their capital cost, as represented by a parameter $\rho$. By tuning $\rho$ from 2\% to 10\%, the test results with SRC metrics (under the setting as $\hat{\Upsilon}=90\%$ and $N-4$) are recorded in Table \ref{tab:test3}.  Note that the network hardening cost and MHERs' total renting cost are represented by LRC and MRC, respectively. With the increase of $\rho$, we observe the monotonous increasing of SRC from 26.89k\$ to 107.20k\$. When $\rho=2\%$ and $4\%$, the MHERs are only option for resilience enhancement. So the LRC remains as zero, while the MRC increases with a higher unit rental cost. When  $\rho$ reaches 6\%, it becomes more economical to use the hybrid strategy of network hardening and MHERs' scheduling. Compared to the instance with $\rho=4\%$, the number of deployed MHERs is reduced from 3 to 2, while distribution lines L1-2 and L3-4 are protected from disastrous events. As a result, the LRC increases from zero to 50.25k\$, and MRC decreases from 53.78k\$ to 29.60k\$. Finally, when $\rho=10\%$, the uncertainty of MHERs' available storage levels could make their deployment less economical than the network hardening. In such instance, it is a better option to only perform proactive hardening for achieving the target of load survivability. Correspondingly, the resilience cost is higher than the rest test cases. 

\begin{table}
	\centering
	\caption{Results with respect to Different Ratio of Renting Cost to Capital Cost ($\hat{\Upsilon}=90\%$ and $N-4$)} \vspace{-5pt}
	\scalebox{0.85}{
		\begin{tabular}{ccccc}
			\toprule
			$\rho$  &  Optimal Planning Scheme  & LRC/k\$ & MRC/k\$   & SRC/k\$ \\
			\midrule		
			2\%  & M1N9, M2N11, M5N7 & 0 & 26.89 & 26.89 \\
			4\%  & M1N9, M2N11, M5N7 & 0 & 53.78 & 53.78 \\	
			6\%  & L1-2, 3-4; M1N9, M2N11  & 50.25 & 29.60 & 79.85 \\
			10\% & L1-2, 2-9, 4-5, 5-6, 6-7, 9-10, 10-11 & 107.20 & 0 & 107.20 \\
			\bottomrule	  
	\end{tabular}}
	\label{tab:test3}
\end{table} 

\subsubsection{Computational Performance of Enhanced N-PC\&CG}
The solution performance of N-PC\&CG algorithm is tested under different levels of damage severity ($k$) and resiliency target ($\hat{\Upsilon}$). The test results of basic  N-PC\&CG algorithm and our enhanced customization are presented in Table \ref{tab:S-PCCG} and Table \ref{tab:S-PCCG-E}. The optimal objective value, the number of outer- and inner-loop iterations as well as the solution time (counted in seconds) are recorded in columns ``OBJ'', ``N\_itr\_Out'', ``N\_itr\_In'' and ``Time/s'', respectively.
We observe that for a fixed $k$, the solution time is monotonously increasing with the growth of $\hat{\Upsilon}$. For example, when $k=4$ and $\hat{\Upsilon}$ changes from 80\% to 95\% (in Table \ref{tab:S-PCCG}), it requires more iterations to converge for both the outer- and inner-loop algorithms, which results in an evident increase of solution time (increasing from 111.2s to 3973.2s, by nearly 36 times). On the other hand, by fixing $\hat{\Upsilon}$ and increasing $k$, it might not lead to a growing number of outer-loop iterations in most cases. The increase of solution time is mainly rendered by more inner-loop iterations. For example, when $\hat{\Upsilon}=95\%$ and $k$ skips from 3 to 5, the outer-loop iterations keeps the same, but 26 additional inner-loop iteration is required by total. In this case, the solution time increases by 65.3\%.

As shown in Table \ref{tab:S-PCCG} and Table \ref{tab:S-PCCG-E}, compared to the basic N-PC\&CG, our enhanced algorithm exhibits significantly improved computational performance. It has reduced the solution time by more than 70\% for all the test cases. Specifically, the average solution time decreases from 1168.2s to 224.8s. For instance, when $k=4$ and $\hat{\Upsilon}=95\%$, the basic N-PC\&CG requires 8 outer iterations and 153 inner iterations to solve the $\mathbf{DDU-RRO}$ formulation. In contrast, the enhanced customization of N-PC\&CG converges in only 5 outer iterations. Also, the iteration number of inner-loop C\&CG drops to 59, which is the key of efficiency improvement. As a result, the total computational time of this test case decreases from 3973.2s to 579.4s (reducing by amazingly 85.4\%). By applying our enhancement strategies, we can observe evidently improved convergence performance of both the outer- and inner-loop algorithms, achieving a strong capacity to obtain exact solutions of $\mathbf{DDU-RRO}$.

\renewcommand\arraystretch{1.00}
\begin{table}[htp]
	\centering
	\caption{Solution Performance of N-PC\&CG Algorithm: 14-Bus Test System} \vspace{-5pt}
	\setlength{\tabcolsep}{3.00mm}
	\scalebox{0.85}{
		\begin{tabular}{cccccc}
			\toprule
			$k$     & $\hat{\Upsilon}$ & SRC/k\$ & N\_Itr\_Out & N\_Itr\_In & Time/s \\
			\midrule
			\multirow{3}{*}{3} & 0.80  & 18.59  & 3     & 3,3,7    & 93.6 \\
			& 0.90  & 22.74  & 3     & 3,3,17    & 317.7  \\
			& 0.95  & 39.49  & 6     & 3,3,20,4,17,26  & 1434.4 \\
			\midrule
			\multirow{3}{*}{4} & 0.80  & 18.59   & 3     & 4,4,7     & 111.2 \\
			& 0.90  & 26.58  & 5    & 4,4,11,17,18    & 841.5  \\
			& 0.95  & 52.49  & 8     & 4,4,20,18,23,23,36,25    & 3973.2  \\
			\midrule
			\multirow{3}{*}{5} & 0.80  & 22.74  & 3     & 3,4,12     & 222.7 \\
			& 0.90  & 39.49  & 5    & 3,4,25,11,16  & 1148.2 \\
			& 0.95  & 57.47  & 6     & 3,4,16,16,23,37   & 2371.2 \\
		\bottomrule
	\end{tabular}}    %
	\label{tab:S-PCCG}%
\end{table} 

\renewcommand\arraystretch{1.00}
\begin{table}[htp]
	\centering	
	\caption{Solution Performance of Enhanced N-PC\&CG Algorithm: 14-Bus Test System} \vspace{-5pt}
	\setlength{\tabcolsep}{4.00mm}
	\scalebox{0.85}{
		\begin{tabular}{cccccc}
			\toprule
			$k$     & $\hat{\Upsilon}$ & SRC/k\$ & N\_Itr\_Out & N\_Itr\_In & Time/s \\
			\midrule
			\multirow{3}{*}{3} & 0.80  & 18.59  & 3     & 3,2,5   & 39.5 \\
			& 0.90  & 22.74  & 3     & 3,2,6    & 47.7  \\
			& 0.95  & 39.49  & 5     & 3,3,11,8,12  & 280.2 \\
			\midrule
			\multirow{3}{*}{4} & 0.80  & 18.59   & 2     & 4,5     & 37.5\\
			& 0.90  & 26.58  & 4    & 4,6,8,9    & 157.9  \\
			& 0.95  & 52.49  & 5     & 4,11,11,17,16    & 579.4 \\
			\midrule
			\multirow{3}{*}{5} & 0.80  & 22.74  & 3     & 3,2,6     & 49.4 \\
			& 0.90  & 39.49  & 4    & 3,2,11,12  & 208.2 \\
			& 0.95  & 57.47  & 6     & 3,2,7,10,18,18   & 623.1 \\
			\bottomrule
	\end{tabular}}%
	\label{tab:S-PCCG-E}%
\end{table} 

\renewcommand\arraystretch{1.00}
\begin{table} [h!]
	\centering
	\caption{Techno-Economical Parameters of MHERs: 56-Bus Real-World System} \vspace{-5pt}
	\setlength{\tabcolsep}{1.00mm}
		\scalebox{0.85}{
			\begin{tabular}{ccccccccc}
				\toprule
				Type &   Number    &   $\overline{H}$  &   $\overline{P}$  & $\overline{Q}$    &  $\xi$    &  $\eta$    &  $c^H$    & $m^{tp}$   \\
				\midrule
				1     & 1/5  & 60kg  & 300kW & 350kVar      & \multirow{4}{*}{38.9kWh/kg} & \multirow{4}{*}{52\%} & 12.60/15.12k\$  & \multirow{3}{*}{1kg/h} \\
				2     & 2/6     & 100kg & 400kW & 450kVar       &       &   & 17.00/20.40k\$  &  \\
				3     & 3/7   & 130kg & 400kW & 450kVar       &       &   & 17.30/20.76k\$  &  \\
				4     & 4/8   & 160kg & 500kW & 500kVar       &       &   & 21.60/25.92k\$  &  \\
				\bottomrule
	\end{tabular}}
	\label{tab:MHER_56}%
\end{table}

\subsection{56-Bus Real-World Distribution Network}
The scalability of the proposed co-planning method is tested on a real-world 56-bus distribution network (SCE 56-bus system \cite{gan2015exact}). It is modified to integrate 7 EH nodes and 8 MHERs. 5 SGs are configured with an identical power rating as 500 kW. All these MHERs are categorized into four types, with their techno-economical parameters presented in Table VII. Four distribution lines L9-19, L6-29, L33-44 and L50-55 are equipped with tie switches and assumed to be invulnerable under disastrous occasions. Given $k=7$ and $\hat{\Upsilon}=85\%$, the co-planning scheme and associated worst-case contingency scenario is given in Fig. 2. Also, Fig. 3 exhibits the post-event scheduling results, i.e., the re-routing of MHERs and network reconfiguration, corresponding to the worst-case damage scenario. Note that the blue shaded regions in two figures contain the vulnerable distribution lines.

\begin{figure*}[!t]
	\centering
	\includegraphics[width=12.5cm]{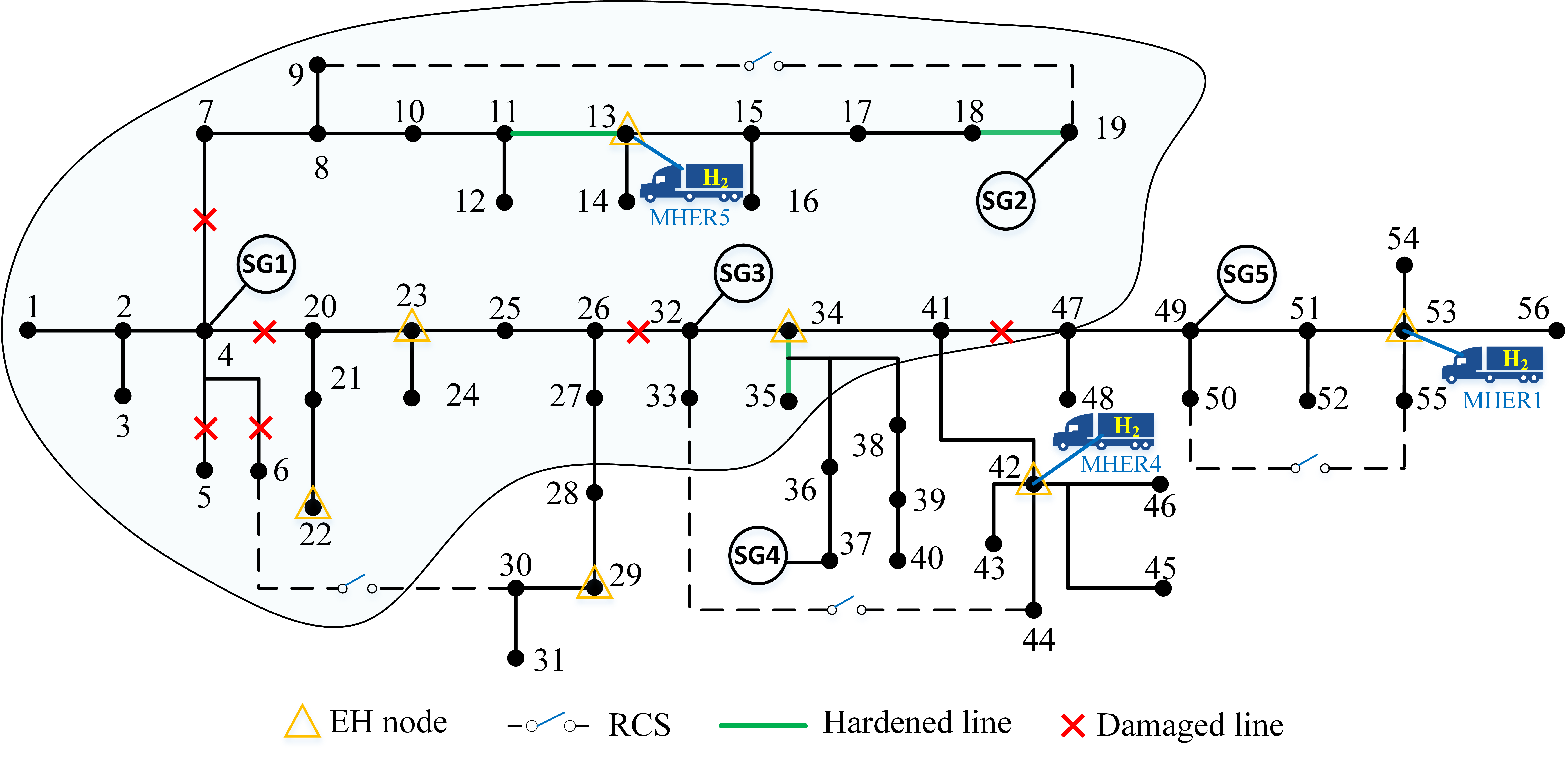}
	\caption{Optimal Resilience Enhancement Plan with $k=6$ and $\hat{\Upsilon}=85\%$: 56-Bus Real-World System }
	\label{fig:Optimal_scheme} 
\end{figure*}

\begin{figure*}[!t]
	\centering
	\includegraphics[width=12.5cm]{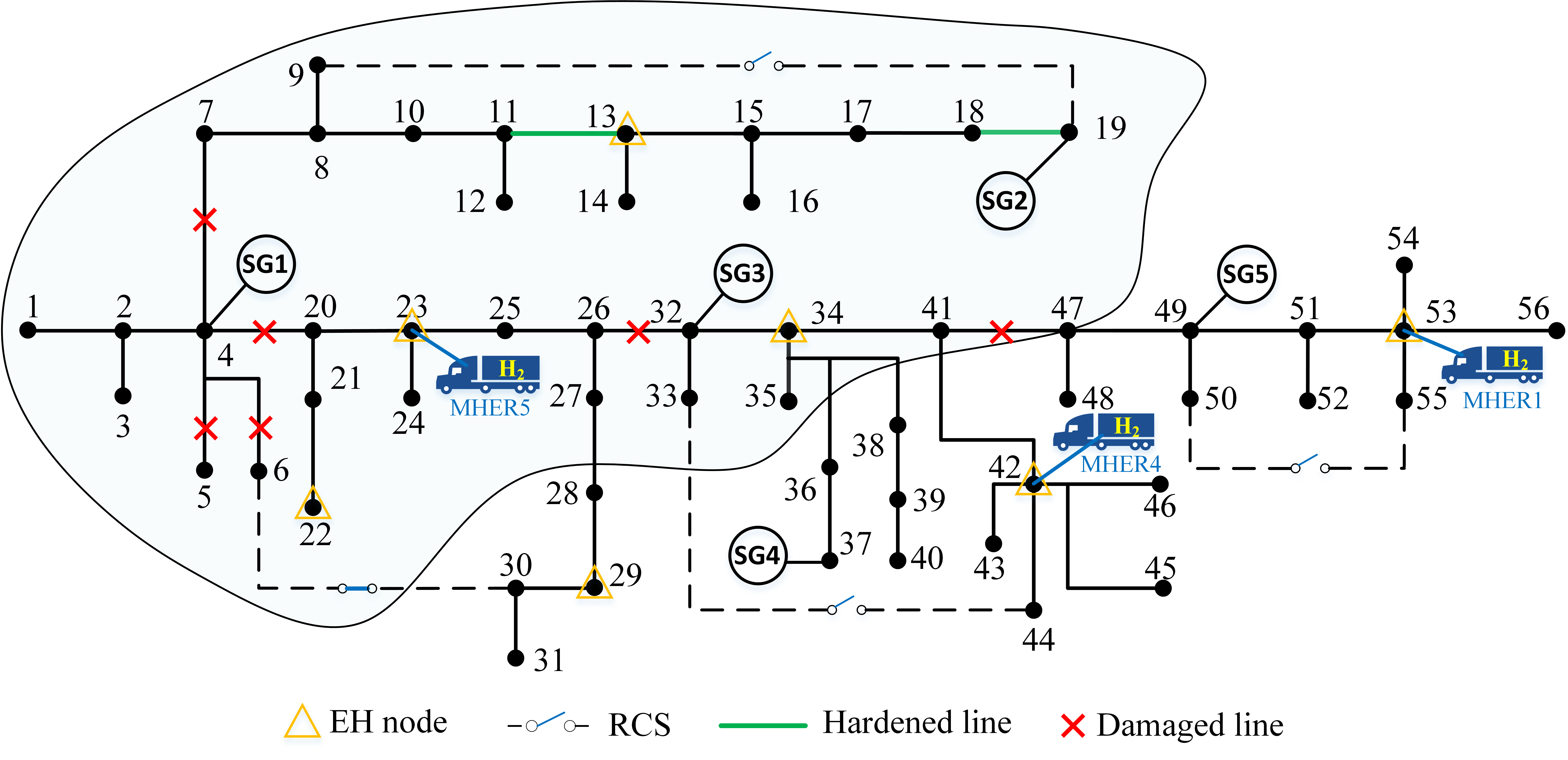}
	\caption{Optimal Re-Routing and Scheduling Results Against Worst-Case Scenario: 56-Bus Real-World System }
	\label{fig:Scheduling_results} 
\end{figure*}

As shown in Fig. \ref{fig:Optimal_scheme}, lines L11-13 and L18-19 are hardened with the proactive allocation of MHER1, 4, 5 at N53, N42 and N13, respectively. Under the worst-case scenario, the network damage scenario is identified as the outages of L4-5, L4-6, L4-7, L4-20, L26-32 and L41-47, partitioning the entire system into 7 parts. The damages of L4-5, L4-6, L4-7 and L4-20 cut off the connection between most nodes and the substation (i.e., N1). As a part of the worst-case scenario, the available storage levels of MHER1/MHER4 and MHER5 reach their lower limit, i.e., 70\% and 90\%, respectively. At the post-event phase, SGs, MHERs and the tie switches are coordinately scheduled to achieve the load survivability target. As exhibited in Fig. \ref{fig:Scheduling_results}, N7-N19 can be restored via SG2 deployed at N19, enabled by the intact protection of lines L11-13 and L18-19. To exploit the flexible emergency resources, MHER5 is dispatched from N13 to N23, with the RCS at L6-29 activated to energize the two islands (i.e., N6 and N20-31). Simultaneously, electric nodes N32-N46 are supplied by SG3 and MHER4, while N47-N56 are supported by SG5 and MHER1. As a consequence, the co-optimization of network hardening and MHERs deployment demonstrates a high cost efficiency, achieving a substantial cost saving by decreasing the SRC from 286.13k\$ to 78.87k\$ (reducing by 72.44\%).

Moreover, the comparative solution tests on enhanced N-PC\&CG and its basic version are conducted under different combinations of $k$ and $\hat{\Upsilon}$. The number of outer-loop iterations and cumulative  inner-loop iterations is recorded in the column ``N\_itr (Out/In)''. The test instances that fail to converge within 12 hours are marked with ``T''. As shown in Table VIII, the basic N-PC\&CG fails to converge within the time limit ``T'' in those test cases where $\hat{\Upsilon}=95\%$. Also, it is rather time consuming for the rest cases. In contrast, our enhanced customization in Section IV-E successfully converges for all the test instances. In comparison to the basic one, the enhanced N-PC\&CG can obtain the optimal resiliency solution using much less time, demonstrating a superior test performance for the real-world power grid. For example, when $k=9$ and $\hat{\Upsilon}=85\%$, the basic N-PC\&CG requires 37 outer-loop iterations with 500 cumulative inner-loop iterations, consuming 14159.3s to converge. By applying the enhanced algorithm, the number of outer- and total inner-loop iterations reduces to 14 and 119, respectively. The corresponding solution time drops to 3578.3s (reducing by 74.7\%). Although for a few very challenging instances (e.g., when $\hat{\Upsilon}=95\%$ and $k=7$ or $9$), the solution time of enhanced N-PC\&CG may be sharply raised to around 4 hours, it is practically acceptable for a long-term planning problem. The aforementioned results have demonstrated the scalable solution capability of our customized decomposition algorithm.

\renewcommand\arraystretch{1.00}
\begin{table}[htp]
	\centering	
	\caption{Solution Performance of Two Types of N-PC\&CG Algorithms: 56-Bus Real-World System} \vspace{-5pt}
	\scalebox{0.9}{
		\begin{tabular}{ccccccc}
			\toprule
			\multirow{2}{*}{$k$} & \multirow{2}{*}{$\hat{\Upsilon}$}  & SRC/k\$ & \multicolumn{2}{c}{Basic N-PC\&CG} & \multicolumn{2}{c}{Enhanced N-PC\&CG}   \\
			&   &    & N\_itr(Out/In) & Time/s   & N\_itr(Out/In) & Time/s   \\
			\midrule
			\multirow{3}{*}{5} & 0.75   & 32.12    & 5/18   & 158.9   & 5/11  & 69.3     \\
			& 0.85   & 46.90   & 4/22    & 256.9   & 4/19 & 226.8          \\
			& 0.95   & 272.86   & /    & /   & 14/149   & 5726.3       \\
			\midrule
			\multirow{3}{*}{7} & 0.75   & 38.60    & 4/23   & 186.2   & 3/11  & 71.6    \\
			& 0.85   & 97.90   & 15/150    & 3726.9   & 11/110  & 2187.5         \\
			& 0.95   & 318.60   & /    & /   & 18/278   & 14974.3       \\
			\midrule
			\multirow{3}{*}{9} & 0.75   & 49.42   & 5/38  & 349.3   & 4/17  & 153.1    \\
			& 0.85   & 140.29   & 37/500    & 14159.3   &14/119  & 3578.3        \\
			& 0.95   & 349.90   & /    & /   & 19/298   & 15900.3      \\
			\bottomrule
	\end{tabular}}
	\label{tab:S-PCCG-56}%
\end{table}

\section{Conclusion}
To support the co-planning of power network hardening and MHERs' scheduling under disastrous occasions, this paper proposes a resilience-constrained RO model with DDU set and discrete recourse structure. The endogenous uncertainties of network contingency and MHERs' available storage levels are analytically represented.  Then, to exactly and efficiently solve this challenging formulation, a nested customization of PC\&CG algorithm is developed. Note that the algorithm is firstly proposed to solve the two-stage RO model with mixed-integer DDU set and discrete recourse decisions. Numerical results on 14-bus test system and 56-bus real world distribution system confirm the effectiveness and scalability of the proposed method. Under different damage levels, it has evidently decreased the capital cost for achieving the same target of load survivability under worst-case scenario. Also, our customized N-PC\&CG algorithm with nontrivial enhancements demonstrates excellent computational performance for addressing practical-scale problems.

\ifCLASSOPTIONcaptionsoff
  \newpage
\fi

\bibliographystyle{IEEEtran}
\bibliography{Ref}

\end{document}